\documentclass{tufte-handout}

\title[A Short History of
Rocks]{\emph{\texttt{\textcolor{myred}{\huge{A Short History of Rocks:}\\
        \LARGE{or, How to Invent Quantum Computing}}}}}

\author[David Wakeham]{\href{mailto:david@torsor.io}{David Wakeham}
 $\diamond$ \href{torsor.io}{Torsor Labs}
}

\date{}

\usepackage{graphicx} 
  \setkeys{Gin}{width=\linewidth,totalheight=\textheight,keepaspectratio}
  \graphicspath{{graphics/}} 
\usepackage{amsmath, amssymb}  
\usepackage{units}    
\usepackage{multicol} 
\usepackage{enumitem}   
\usepackage{hyperref}
\setcitestyle{numbers}
\usepackage{cjhebrew}

\usepackage{bbm}
\usepackage{euler}

\definecolor{timelifeblue}{RGB}{51, 102, 204}
\definecolor{myred}{RGB}{193, 45, 45}
\definecolor{mylinks}{RGB}{20, 20, 20}

\usepackage{listings}
\definecolor{codegreen}{rgb}{0,0.4,0.3}
\definecolor{codegray}{rgb}{0.5,0.5,0.5}
\definecolor{codepurple}{rgb}{0.58,0,0.82}
\definecolor{backcolour}{rgb}{0.95,0.95,0.92}

\usepackage{plex-mono}

\usepackage[tabular,lining]{sourcesanspro}

\begin{document}

\maketitle
\vspace{-15pt}

\par\noindent\rule{0.95\textwidth}{0.4pt}
\vspace{10pt}

\begin{center}
  \textsc{abstract}
\end{center}
This essay gives a short, informal account of the development of
digital logic from the Pleistocene to the Manhattan Project, the introduction of
reversible circuits, and Richard Feynman's allied proposal for quantum computing. We
argue that Feynman's state-based analogy is not the only way to arrive
at quantum computing, nor indeed the simplest.
To illustrate, we imagine an alternate timeline in which John von
Neumann skipped Operation Crossroads to debug a military computer,
got tickled by the problem,
and discovered a completely different picture of quantum computing---in 1946.

Feynman suggested we ``quantize'' state, and turn classically
reversible circuits into quantum reversible, unitary ones.
In contrast, we speculate that von Neumann, with his
background in functional analysis and quantum logic, would seek to
``quantize'' the operators of Boolean algebra, and with tools made available in
1946 could successfully do so.
This leads to a simpler, more flexible circuit calculus and beautiful
parallels to classical logic, as we detail in a forthcoming companion paper.
\begin{abstract}
\noindent

\end{abstract}

\vfill 
\hspace{-25pt} \includegraphics[width=0.25\columnwidth]{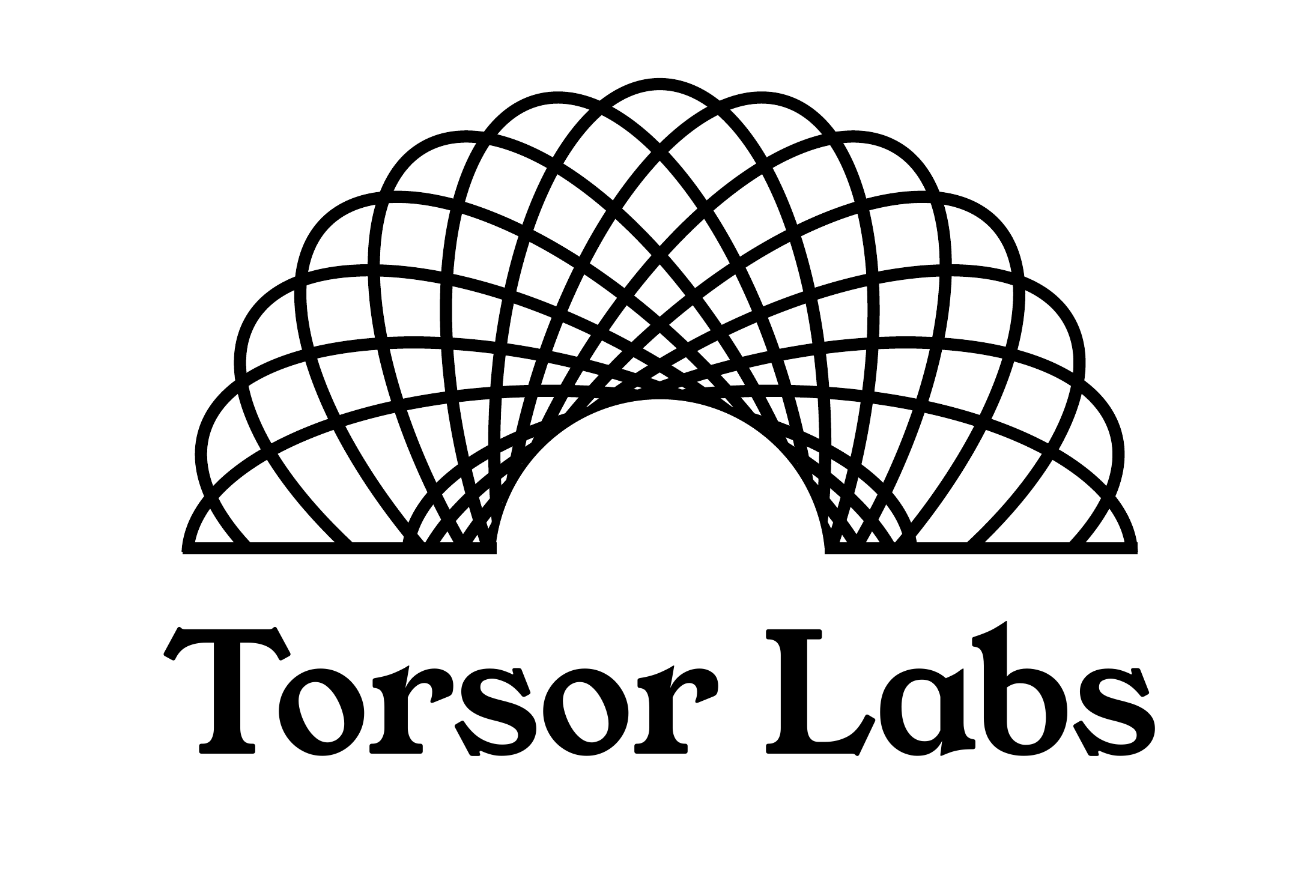} \\
\vspace{-10pt}
\noindent\hspace{0pt}\textsf{YAW}-\<'>${}_\texttt{0}$-2-25

\newpage

\section*{\LARGE{\emph{0} \hspace{5pt} Overview}}

\noindent This essay gives a short history of human-rock interactions, arguing that, far from tricking rocks into doing math, they
tricked us; or rather, the trickery is mutual and ongoing. We support this thesis by example, progressing through number
systems, binary mysticism, Boolean logic, digital circuits, and large-scale programmable
architectures, drawing attention at each juncture to the collaborative
role played by our igneous friends. We conclude the first half with quantum
computing, in some sense the apotheosis of this mutual
trickery.

History abhors a linear narrative; in our case, the linear narrative is
undone by the Manhattan Project, which led to a long chill between the
physicists who worked on the bomb---including Richard Feynmann---and
the ductile rocks that helped turn an equation as
beautiful as $E = mc^2$ into Little Boy and Fat Man.
The physicists left rocks behind and turned to theory. Although the rocks tried
to teach us quantum, for a long time, we didn't listen; when eventually,
reluctantly, we did, the transmission was garbled.

In the second half of this essay, we try to clarify what the rocks
might have meant, using the clues scattered throughout history and
some generous counterfactual license.
It may be that, 
in a branch of the wavefunction not so far from this one, John von Neumann
invented quantum computing $35$ years before Feynman, using
functional analysis to generalize Shannon's algebra of circuits.
We give a fuller development of this formalism elsewhere, but hope
that in the mean time, the reader is a little more attuned to the
whispering presence of the inanimate world.

\begin{flushleft}
  \textsc{acknowledgments}
\end{flushleft}

  \noindent Thanks to 
  Jonah Berean-Dutcher, Pompey Leung, and Abhisek Sahu for feedback on the draft, Ruth
  Wakeham for rediscovering the Life Nature Library with me, and
 Jon Male and Clara Weill for encouraging me to get out and talk to
 rocks.

\vspace{5pt}
 
\par\noindent\rule{0.7\textwidth}{0.4pt}

\vspace{-30pt}

\tableofcontents
 
\newpage

\vspace*{10pt}

\section[\emph{1} \hspace{5pt} The other calculus]{\LARGE{\emph{1}
    \hspace{5pt} The other calculus}}\label{sec:calculus}

\marginnote{
  \begin{center}
    \includegraphics[width=0.9\linewidth]{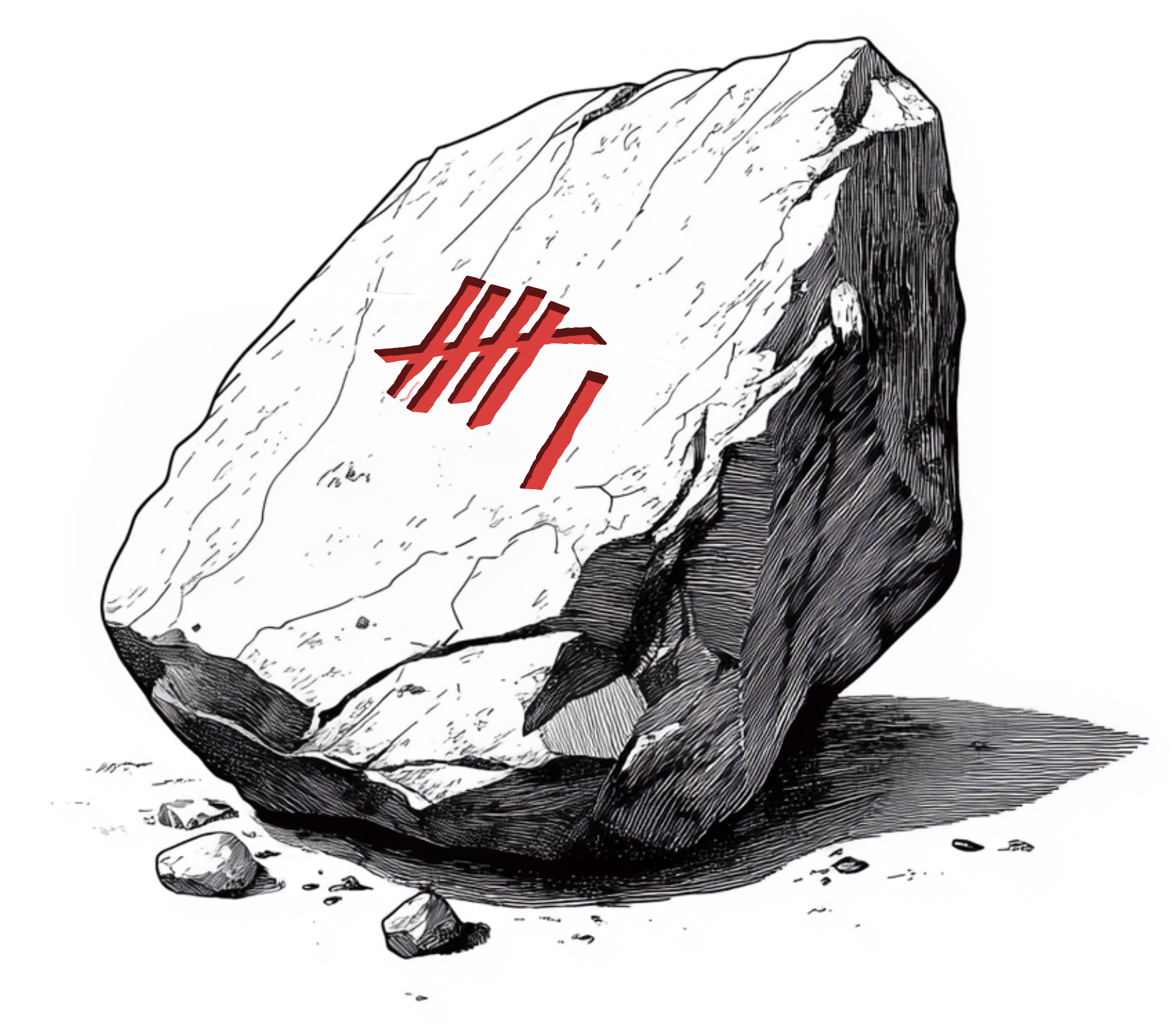}
  \end{center}
  \vspace{-10pt}
  \emph{Unary programming in silico, aka tallying.
    Tallies exist because we ran out of fingers.
  }
}
\noindent The development of mathematics and technology is deeply tied to our interaction with rocks.
During the Pleistocene, humans discovered the number $1$ and
began to count in unary on their fingers, with pebbles, or by
repetitive scratches onto stone or slate.
We can think of these techniques as a unary programming language
(a formal system for representing algorithms) and the rocks as an
early form of computer (a means of carrying those algorithms out).

The transition from hands to tally marks is instructive.
Tallies were invented because we ran out of fingers; each round of
five marks effectively supplies an extra hand.
In fact, the diagonal slash for grouping tallies is an
\emph{abstraction} of the hand, an abstraction forced by computational necessity.
As civilization grew, numbers grew with it, and our
number systems---with associated abstractions and algorithms---had to keep up.
\marginnote{
  \begin{center}
    \hspace{-5pt}\includegraphics[width=0.7\linewidth]{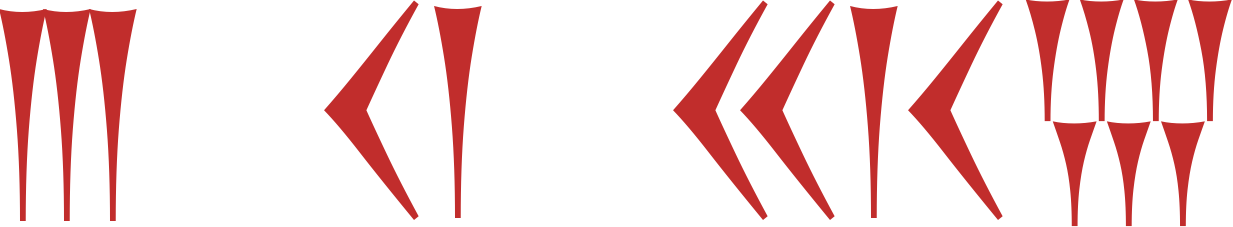}
  \end{center}
  \vspace{-3pt}
  \emph{
    Ontogeny recapitulates phylogeny in Babylonian numerals. 
    \emph{Left.} A unary tally for $3$. \emph{Middle.}
    Sign-value representation of $11$. \emph{Right.} $1277$ in place-value notation.
  }
  }Tally marks became sign-value systems (like Roman numerals, or Sumerian/Babylonian cuneiform
numerals below $60$) and then place-value systems (like Babylonian sexagesimals). The
sequence of hierarchizing macros in between was not so different from
the slash which replaced the hand.


During this evolution, stones were supplanted by more portable
computers like clay, papyrus and parchment.
\marginnote{
  \begin{center}
    \includegraphics[width=0.75\linewidth]{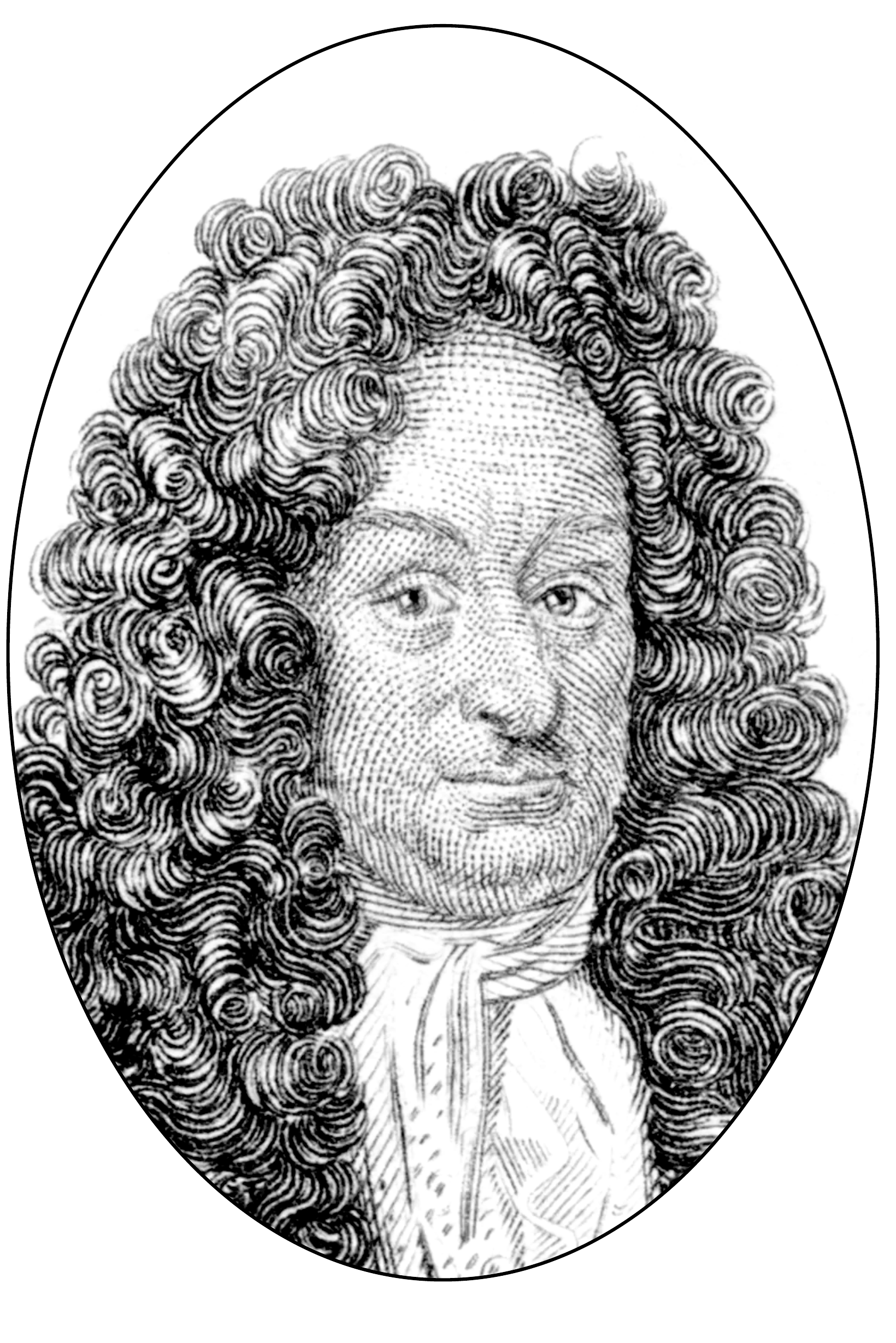}
  \end{center}
  \vspace{-10pt}
  \emph{Gottfried Leibniz (1646--1716). Mathematician, Sinophile, and
    peruke enthusiast.
  }
}The rocks---our steadfast, inanimate friends---would eventually make a 
comeback, but not before the wheel of fortune had revolved a few
times.
Place value systems require a way to indicate place.
While the
Babylonians sidestepped the problem, hoping context or an empty space
would make it clear, Indian mathematicians introduced the
symbol ``$\bullet$'' as a (literal) placeholder. This eventually morphed
into the digit ``$0$''. In contrast to lumbering bases like sixty (probably chosen
for its compositeness) and ten (once again, replacing hands), it
was now possible to build a number system from $0$ and
$1$ alone. 

It took people a while to catch on.
A full 1500 years after $0$ first appeared, German
polymath \textsc{Gottfried Wilhelm Leibniz} began to
noodle around with binary arithmetic.
Leibniz was a Sinophile (among many other things) and learned
that the ``hexagrams'' of the Chinese \emph{Book of Changes}
corresponded to
binary labels.\footnote{To be clear, these did not constitute a place value system.} 
He took this parallel---across a vast interval of time, language,
and culture---as proof of the underlying universality of human
thought.

This universality would become one of his central preoccupations.
First, it suggested the possibility of a universal language or \emph{characteristica universalis} (``general
characteristic''), which he envisioned as a fantastically expressive pictorial script ``by which all concepts
and things [could] be put into beautiful order,''\sidenote{``On the
  General Characteristic'' (1679).}
but which would be precise enough to reason about
mathematically. To actually perform this reasoning, Leibniz introduced
another powerful abstraction: the \emph{calculus
  ratiocinator} (``reasoning calculator''), a  ``general algebra in
which all truths of reason would be reduced to a kind of
calculus.''\sidenote{Letter to Nicolas Remond (1714).}\marginnote{
  \begin{center}
    \includegraphics[width=0.75\linewidth]{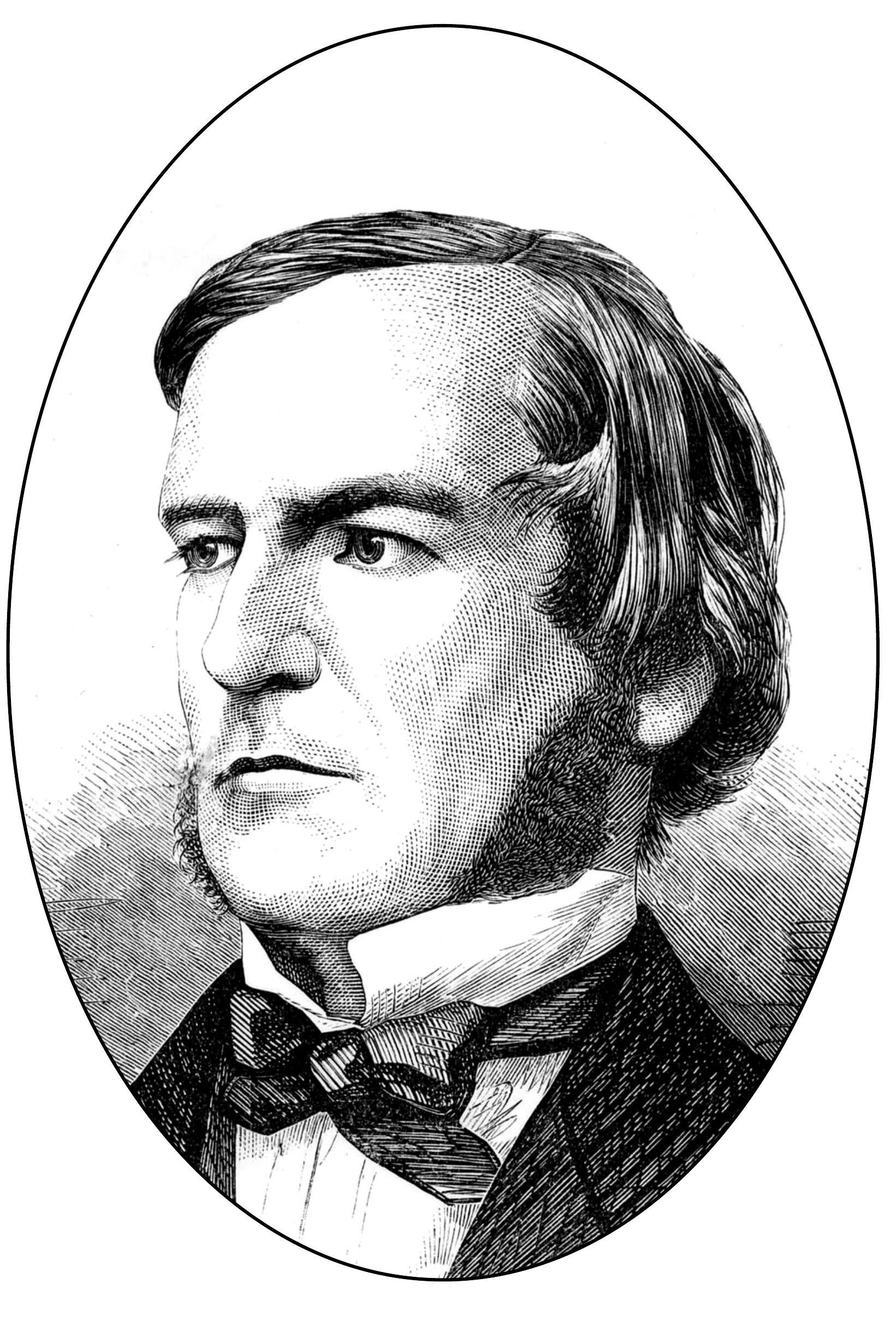}
  \end{center}
  \vspace{-10pt}
  \emph{George Boole (1815--1864). 
    Son of a Lincolnshire cobbler, heir of Leibniz.
  }
}
Though less well-known than Leibniz's work on infinitesimal calculus, the
\emph{characteristica} and the \emph{ratiocinator} are no less
important. They laid the conceptual foundations of modern digital programming
(\emph{characteristica}) and computation (\emph{ratiocinator}).

He didn't live to see either utopian project realized.
But a century after Leibniz's death, the wife of a downtrodden Lincolnshire cobbler
gave birth to a son.
The cobbler was more enthusiastic about science than shoes, 
and his child---\textsc{George Boole}---would leave school early to
pick up the slack, 
teach himself mathematics in his spare time, and blaze the trail to a professorship at
Queen's University. He would also build the ``general algebra'' that Leibniz had
dreamt of, and from the binary components of the \emph{characteristica},
no less.

Boole's crucial insight was that logic could be
\emph{algebraized}.
If we identify $1$ with
\texttt{True} and $0$ with
\texttt{False}, then basic logical operations like \texttt{AND} ($\wedge$),
\texttt{OR} ($\vee$),
and \texttt{NOT} ($\neg$) become algebraic:
\[
  x \wedge y = x\cdot y, \quad x \,\vee \, y = x + y - xy, \quad \neg x = 1 - x,
\]
\marginnote{
  \begin{center}
    \includegraphics[width=0.55\linewidth]{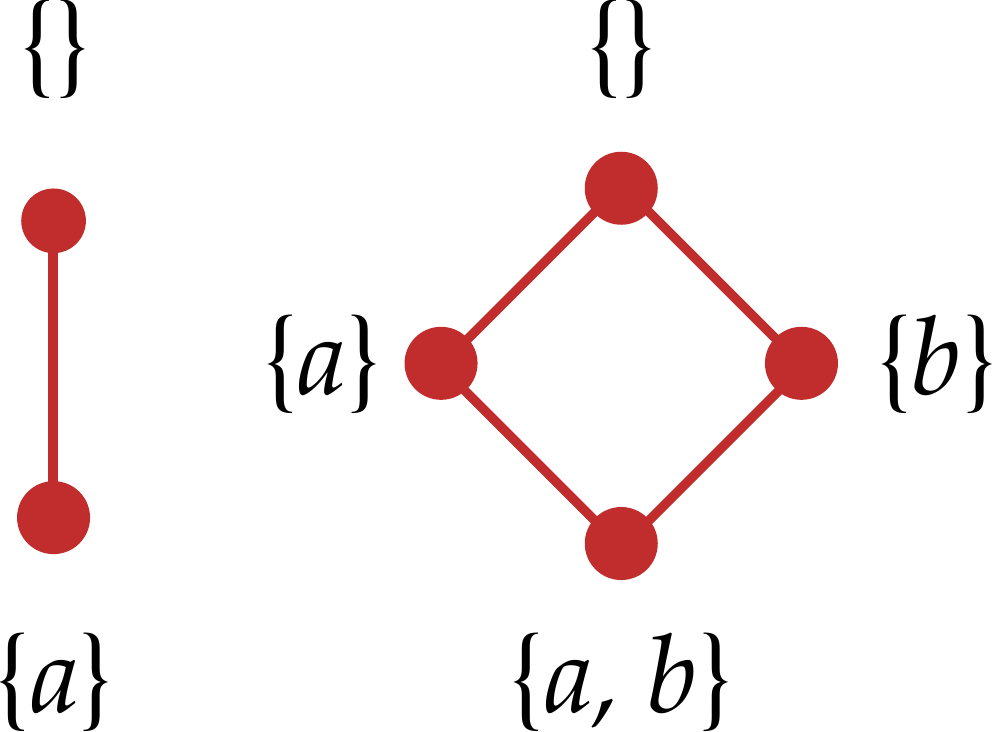}
  \end{center}
  \vspace{0pt}
  \emph{A concrete realization of Boolean algebra using
    sets. Logical operations correspond to set operations: union ($\vee$),
    intersection ($\wedge$), and complement ($\neg$).
  } 
}where $x, y$ are variables representing propositions,
and the operations on the
right are ordinary arithmetic.
These observations are surprisingly powerful. For instance, the
algebraic fact that
\[
  1 - xy = (1 - x) + (1 - y) - (1 - x) (1 - y)
\]
is equivalent to \textsc{De Morgan's Law}:
\[
  \neg (x \wedge y) = \neg x \, \vee \, \neg y.
\]
This structure, called a \textsc{Boolean algebra}, indeed reduces the
``truths of reason'' to a remarkably simple calculus.\sidenote{See \emph{The Mathematical Analysis of Logic}
  (1847) and \emph{An Investigation of the Laws of Thought} (1854).}

Let's recap. We started counting $1$s, ran out of fingers, drew on rocks,
ran out of room, invented sign value, place value, then $0$, and
ported it all to lighter computers. That was good for a while, until a
strange  little man with a wig started counting in $1$s \emph{and} $0$s, noticed the same in a
2500-year old Chinese divination manual, figured that made a good case for a
universal calculus, which a shoemaker's son cobbled together, from $1$s
and $0$s, 150 years later. The scene was now set for the return of the
rocks.

\section[\emph{2} \hspace{5pt} Computing at scale]{\LARGE{\emph{2}
    \hspace{5pt} Computing at scale}}\label{sec:thinking}

\marginnote{
  \begin{center}
    \includegraphics[width=0.75\linewidth]{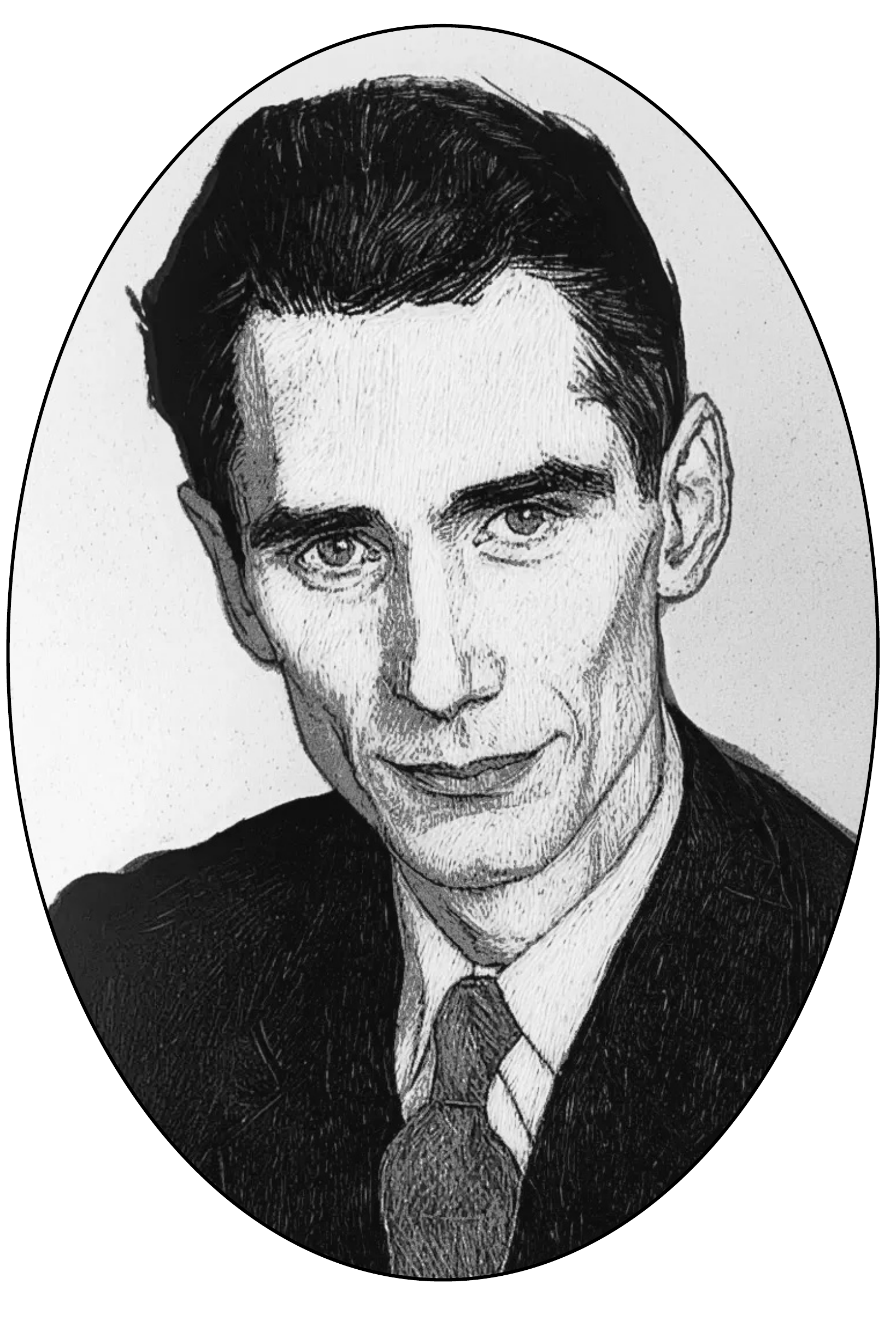}
  \end{center}
  \vspace{-10pt}
  \emph{Claude Shannon (1916--2001).
    The quiet magician who tricked rocks into thinking.
  }
}

\textsc{Claude Elwood Shannon} encountered the work of George Boole as an
undergraduate doing dual degrees in mathematics and electrical
engineering.
Born in the dusty
crossroads of Gaylord, Michigan, Shannon enjoyed puzzles, games, and
taking apart old machinery only to reassemble it in surprising new
ways.
During his masters at MIT, he was tasked with studying the
``differential analyzer''. The
analyzer---a steampunk vision of electromechanical relays and ad hoc
circuitry---was built to solve differential equations (ironically one of Boole's
main interests as a mathematician) and pioneered by the brother of Lord Kelvin (one of Boole's close friends).
Shannon's fitting tribute to Boole was to deconstruct the messy circuitry of
the analyzer and
systematically resynthesize it with Boolean algebra.
In the process, he invented modern digital circuitry.

After a postdoc at the Institute for Advanced Study (IAS)---where Einstein called him ``a brilliant, brilliant
man''\sidenote{\emph{The Idea Factory} (2013), Jon Gertner.}---Shannon
hopped over the river from New Jersey to Bell Labs, then based in Manhattan.
A gaunt, courteous wizard who kept to himself (though he
sometimes roamed the Labs by unicycle), he
would abundantly justify Einstein's praise.
Shannon went on to single-handedly create
modern cryptography, information and communication
theory,\sidenote{See ``A Mathematical Theory of Cryptography'' (1945),
  ``A Mathematical Theory of Communication'' (1948).}
establishing that any contentful message could be
converted into a stream of $1$s and $0$s to be processed on the
Boolean circuits he had devised as a graduate student.
\marginnote{
    \vspace{-10pt}
  \begin{center}
    \includegraphics[width=0.9\linewidth]{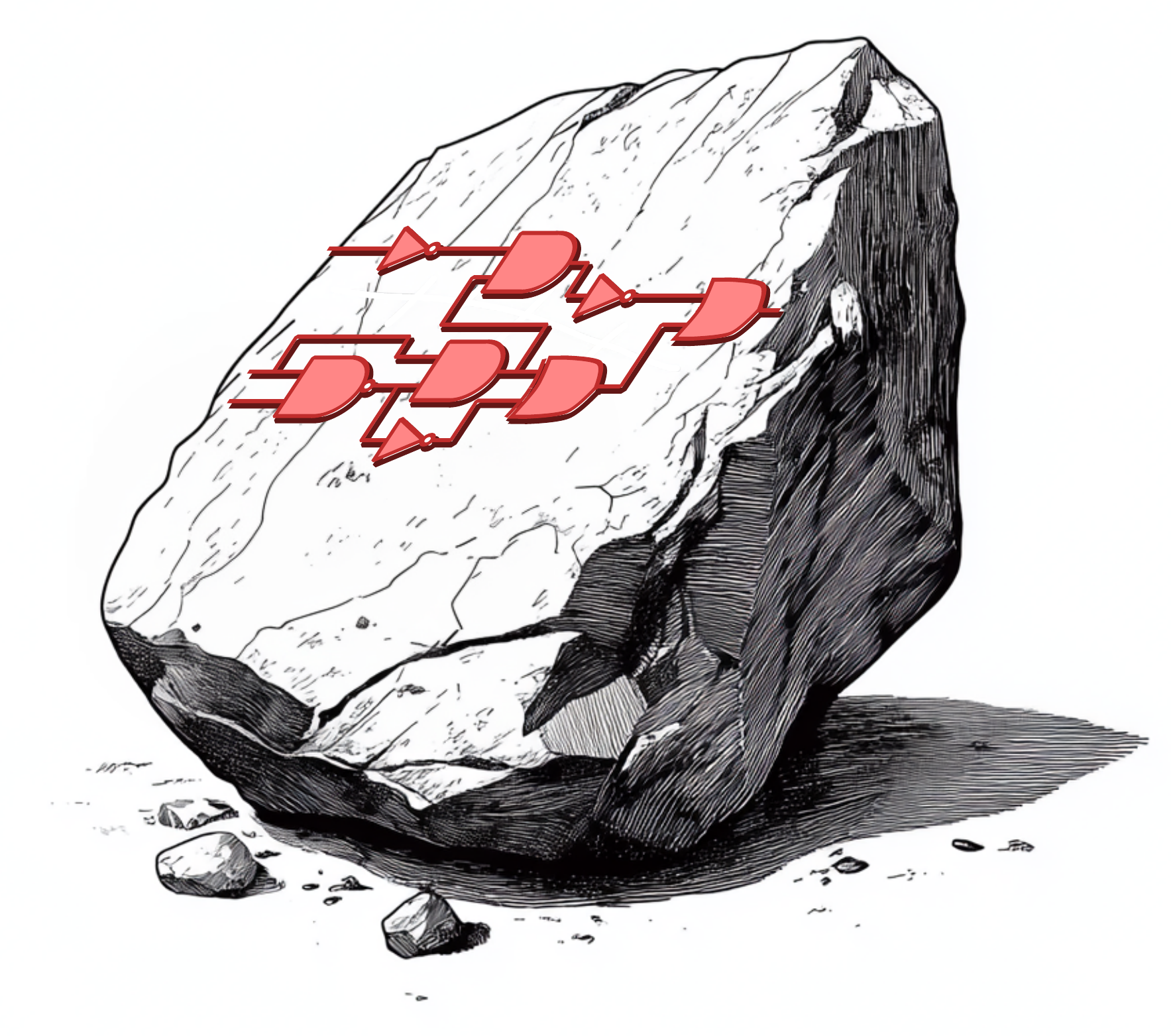}
  \end{center}
  \vspace{-10pt}
  \emph{Binary programming in silico, aka digital circuits.
    These exist because we ran out of fingers, and found a
    symbol for ``ran out''.
  }
}This created a bridge from language to computation,
or in Leibnizian terms, the \emph{characteristica universalis} to the \emph{calculus
  ratiocinator}.
Shannon's ``A Symbolic Analysis of Relay and Switching Circuits''
may be the greatest masters thesis ever written, but 
the sequel was better than the original.

A tally is a line scratched in rock; a circuit is lines drawn in metal.
Though \emph{ratiocinators} would grow ever larger and more
sophisticated, becoming the smartphones and laptops and high-performance GPUs we have today,
many layers of abstraction down is a ``general algebra'' of $1$s and
$0$s, playing across the metal in bursts of current.
It took a few thousand years, but we tricked rocks into doing binary.

While at the IAS, Shannon crossed paths with \textsc{John von
  Neumann}, the Hungarian-American mathematician and youngest faculty
member at the Institute.
The story goes that, by 1940, Shannon had
already struck upon his famous formula
\[
  S[p] = -\sum_{i=1}^n p_i \log_2 p_i = \mathbbm{E}_p[-\log_2 p]
\]
for the amount of information contained in a probability distribution
$p = (p_i)$. But he struggled with the name, hesitating between ``information'' and ``uncertainty''. Von
Neumann offered a third option:\sidenote{``Energy and information''
  (1971), Myron Tribus and Edward McIrving.}
\begin{quotation}
  You should call it \emph{entropy}, for two reasons. 
  In the first place your uncertainty function has been used in
  statistical mechanics under that name. In the second place, and more
  importantly, no one knows what entropy really is, so in a debate you
  will always have the advantage.\sidenote{``Torsor'' may have been
    chosen according to the same guiding principle.}
\end{quotation}
It's unlikely Shannon needed the advantage, but the name
stuck.
\marginnote{
  \vspace{-0pt}
  \begin{center}
    \includegraphics[width=0.75\linewidth]{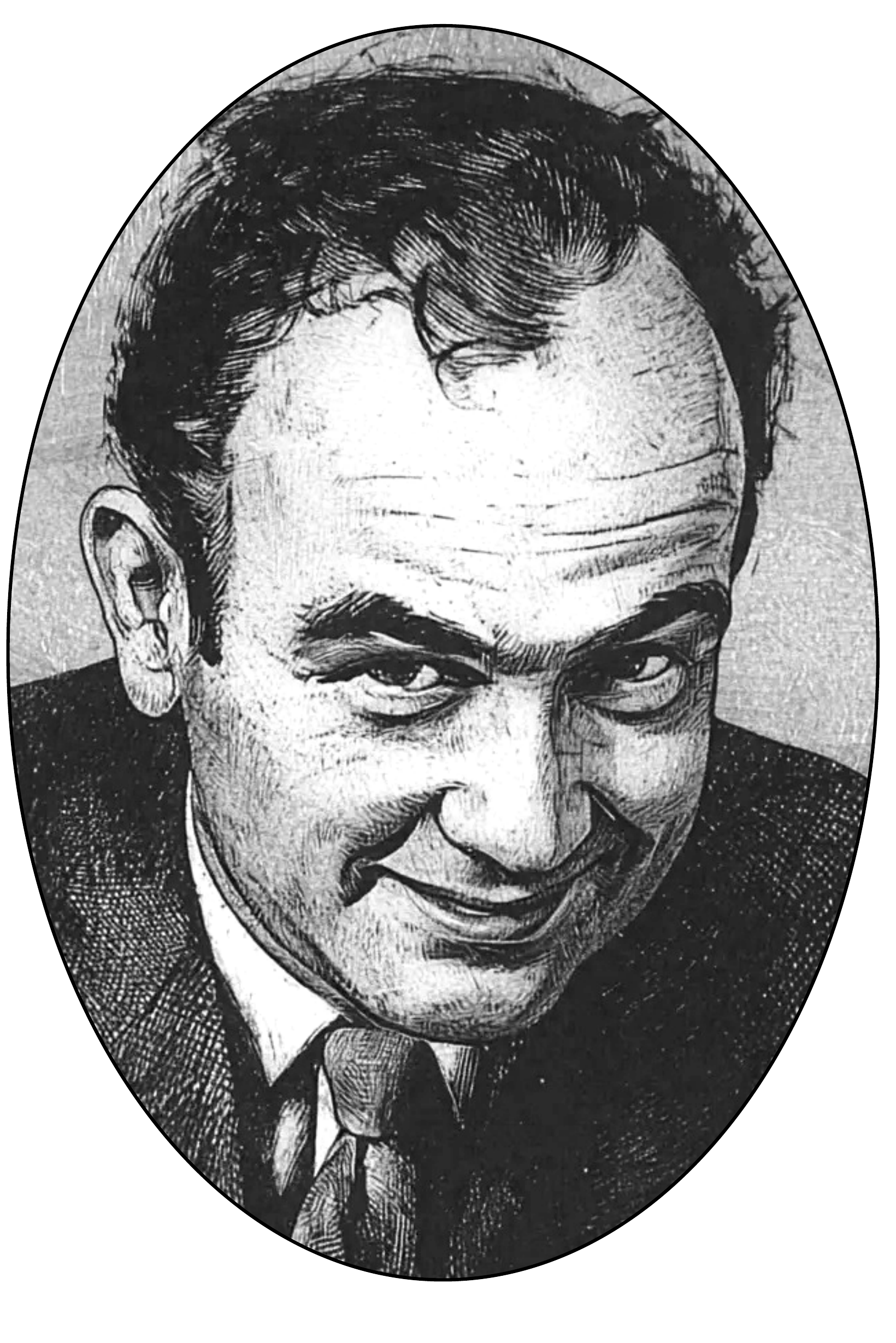}
  \end{center}
  \vspace{-10pt}
  \emph{Here's Johnny! (1903--1957)
    A genius with a strange urge to blow things up.
    \vspace{10pt}
  }
  }

Von Neumann was, in many ways, the opposite of Shannon. Boisterous,
earthy, outgoing, von Neumann 
hailed from Budapest, the glittering capital of the Austro-Hungarian
empire; Shannon was a
wallflower from the backwaters of the Midwest. During WWII, Shannon
stayed at Bell Labs, hoping to quietly avoid the draft; von Neumann
signed up immediately and was rejected due to age, not zeal. And where Shannon was
thorough, methodical and focused, sometimes letting a problem steep for years,
von Neumann was broad and almost inhumanly quick, cutting a dazzling
    swathe through 20th century mathematics both pure and applied. As Hans Bethe wrote\sidenote{``Passing
  of a Great Mind'' (1957), Clay Blair Jr.}
\begin{quotation}
  I have sometimes wondered whether a brain like von Neumann's does
  not indicate a species superior to that of man.
\end{quotation}
Shannon called him the smartest man he had ever
met. But despite their differences, the two shared a yen for applied
problems (both had degrees in engineering) and would, increasingly,
spend their time thinking about computers, conduits of the entropy that
Shannon had fathered and von Neumann baptized.

While Shannon laid low, 
the war took von
Neumann to Los Alamos, where he worked with characteristic vigour on
the science of blowing things up.
\marginnote{
  \vspace{-10pt}
  \begin{center}
    \includegraphics[width=0.95\linewidth]{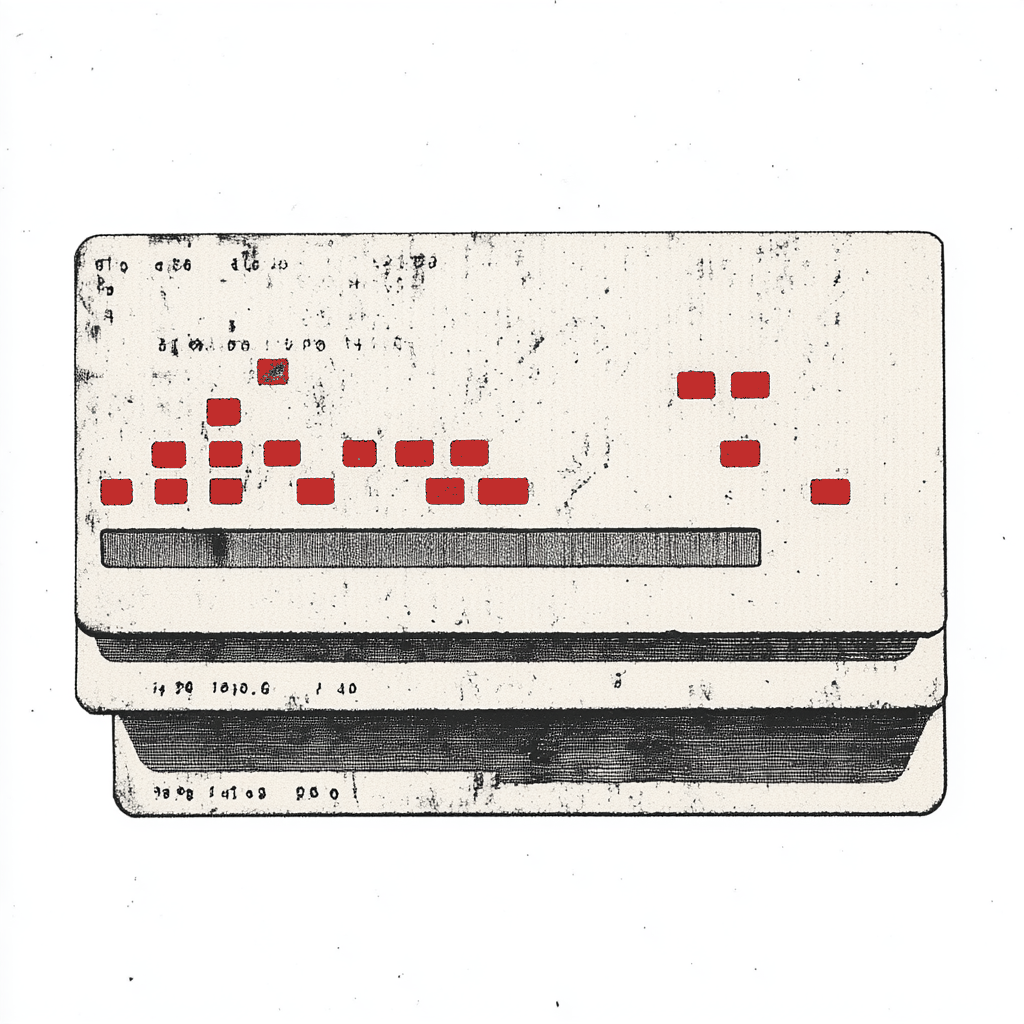}
  \end{center}
  \vspace{-27pt}
  \emph{ENIAC punch cards. As tallies led to sign value, the inconvenience of punch cards
    led to the ``von Neumann'' architecture.
  }
}Part of this science was numerical, and involved a
new toy from the US Ballistic
Research Laboratory: the \textsc{Electronic Numerical
  Integrator and Computer (ENIAC)}. ENIAC was a little like the
differential analyzer Shannon had studied, but larger, faster, and
most importantly, \emph{programmable}.
You could run different programs simply by swapping out punch cards.
Von Neumann would develop some of the first, very primitive, programming languages in
order to tell ENIAC how to run thermonuclear simulations.

Programming on punch cards is a bit like counting on fingers; it only
works for small problems.
The architects of ENIAC (\textsc{John Mauchly} and \textsc{J. Presper Eckert}) realized
they needed a way to store programs and data, and Eckert
invented a clever memory
unit based on pinging signals through mercury. This was one of a
number of innovations bundled into ENIAC's successor, the \textsc{Electronic Discrete Variable Automatic
  Computer (EDVAC)}, on which Mauchly and Eckert gave lectures in 1945.
Von Neumann took polished, comprehensive notes, peppered with original insights,
which an incautious colleague began to circulate; the notes went viral, quashing Mauchly and Eckert's
patent claims and leading to the permanent misattribution ``von Neumann
architecture'' for EDVAC's design scheme. Von Neumann's reputation
preceded not only himself, but his colleagues as well.

Whatever the precise division of credit, von Neumann was central
to the early history of computing at scale,
creating the first protocols for talking to ENIAC/EDVAC, its first
applications, its first bespoke algorithms (merge sort and Monte
Carlo approximation with Stan Ulam), and aspects of the first
integrated, stored-program
architecture.
After the war, he would increasingly focus on methods
for large-scale numerical and scientific computing, including the
first climate-modeling software, run on
ENIAC.\sidenote{``Numerical Integration of the Barotropic
    Vorticity Equation'' (1950), with Jule Charney and Ragnar
  Fjørtoft.}
We can only wonder what else his marvelous organic brain might have
achieved in concert with the electronic brains at his disposal.
Von Neumann died from cancer in 1957, probably caused by wartime
radiation exposure.
He left his mark on the bomb; it left its mark on him.

\marginnote{
  \begin{center}
    \includegraphics[width=0.75\linewidth]{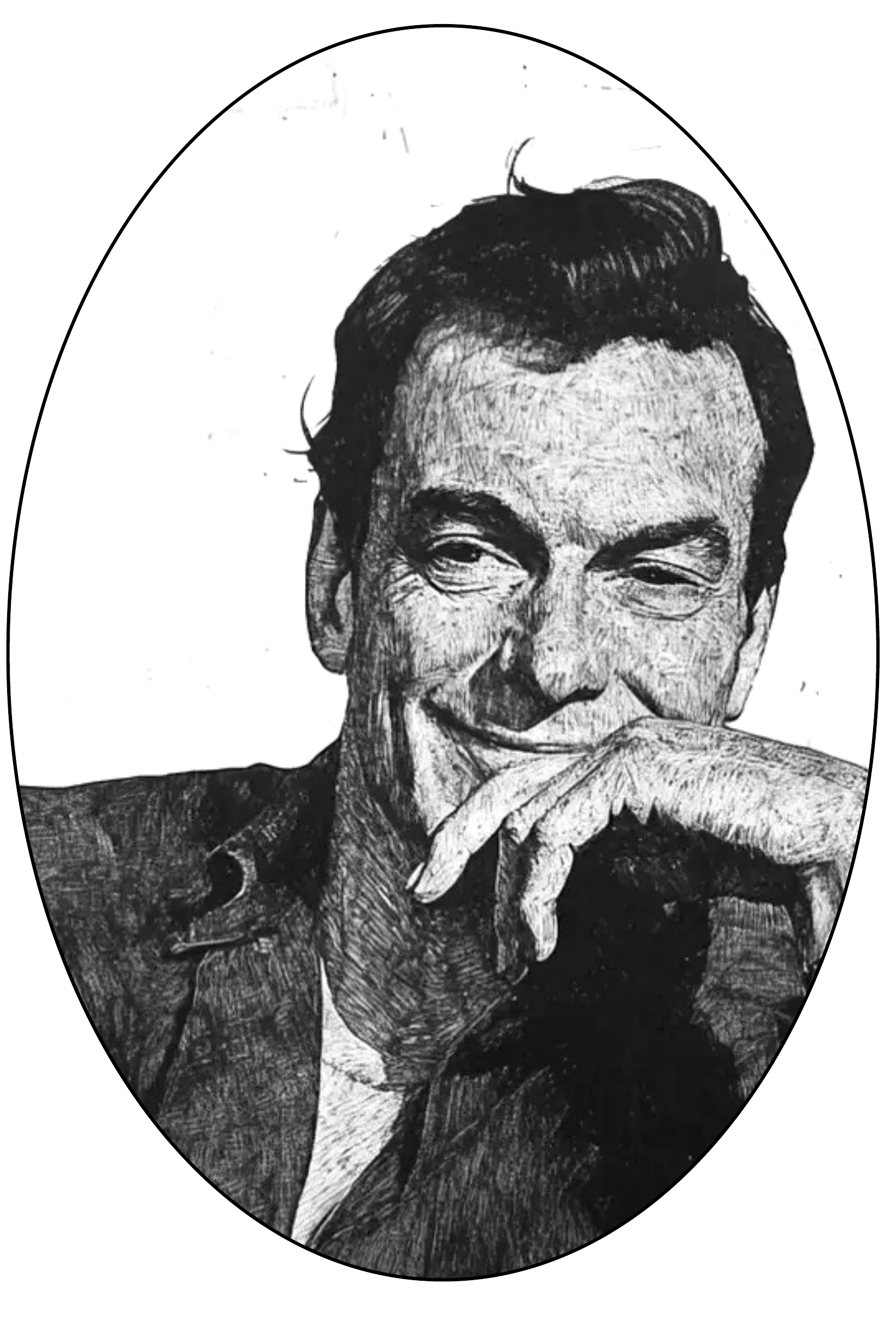}
  \end{center}
  \vspace{-10pt}
  \emph{Richard P. Feynman (1918--1988). The folk hero of the very small.
    \vspace{10pt}
  }
}
The Manhattan Project also left subtler marks.
Von Neumann's assistant on the ENIAC simulations---handling punch card
operations---was a bright young theoretician called \textsc{Richard Feynman}.
Feynman would later win a Nobel Prize for his work on quantum
electrodynamics, and become legendary for his originality, intuition, and
goofy, homespun charm.
In his autobiography,\sidenote{\emph{Surely You're Joking,
    Mr. Feynman!} (1985).} he makes his time at the Manhattan Project
sound like a sequence of wise-cracking, safe-cracking hijinks.
But as historian Cathryn Carson soberly observes:\sidenote{``An Eden after the
  Fall'' (1993).}
\begin{quotation}
  The reality was somewhat grimmer: the
  coded letters, for instance, were to his wife, his highschool
  sweetheart, dying of tuberculosis in a cheap sanitorium outside
  Albuquerque. The real lessons Feynman learned at Los Alamos [were] how
  to hide his feelings behind a brash facade and how to excise
  unwelcome memories.
\end{quotation}
Feynman was perhaps less happy to estimate a death
toll, or the optimal height to detonate a bomb, than the hawkish von
Neumann.

Feynman may have distanced himself emotionally by becoming a ``curious
character,'' the bongo-playing beatnik and hero of every
anecdote. But he
also distanced himself scientifically.
\marginnote{
  \begin{center}
    \includegraphics[width=0.95\linewidth]{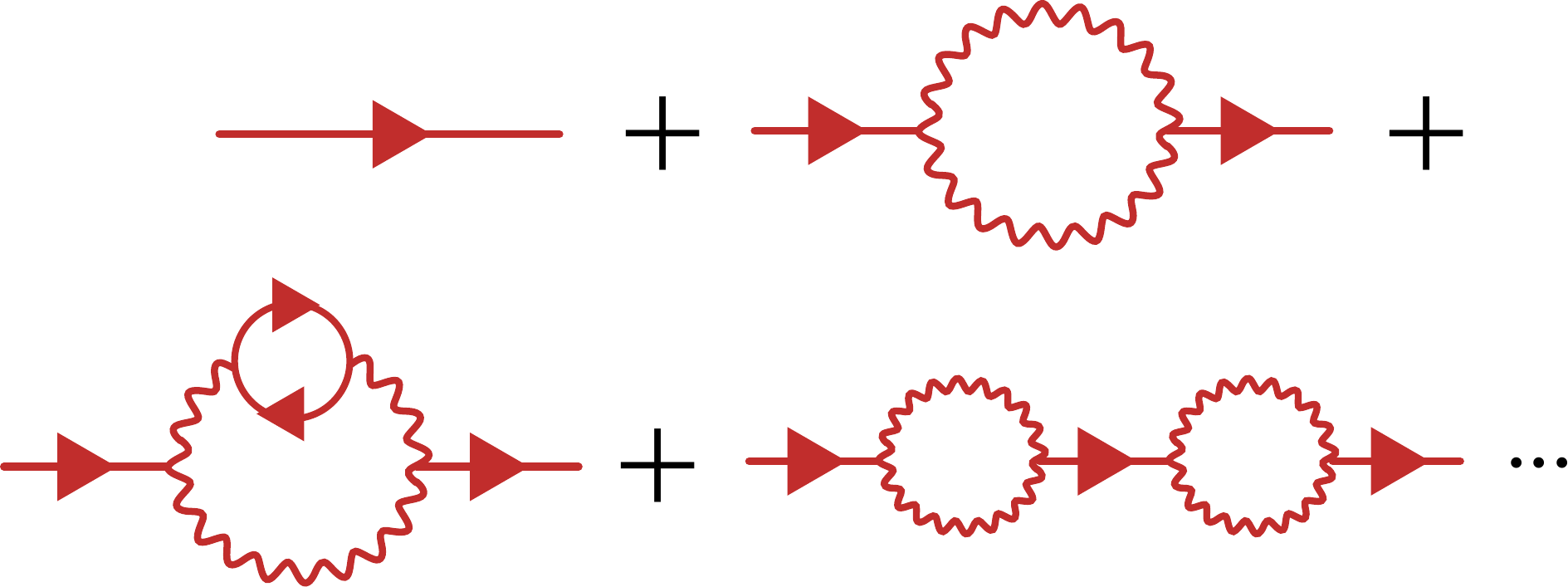}
  \end{center}
  \vspace{-3pt}
  \emph{Feynman diagrams for an electron minding its own business (aka
    the electron propagator). The wiggles are virtual photons.
    \vspace{10pt}
  }
}In contrast to von Neumann, the
champion of large-scale computation, Feynman would turn to the physics of
the very small.
His Nobel Prize-winning work
made the leap from the quantum mechanics of point-like particles to
spatially extended objects called \emph{fields}, and thereby helped establish the
framework of \emph{quantum field theory}.
Perhaps his most famous contribution was
a graphical technique called \emph{Feynman diagrams} for approximating the
probability that one set of particles will collide and transform into
another set.

Feynman's gift for the very small was not just theoretical.
In 1959, he gave a prescient lecture to the American Physical Society called ``There's Plenty of Room at
the Bottom'', with the general theme of tricking tiny
rocks into doing our bidding. This more or less inspired the 
field of nanotechnology. One thought experiment---adapted, in fact, from von
Neumann---was \emph{scaled replication}, where a hierarchy of ever smaller
robot hands is used to eventually build at the nanoscale. The lecture wasn't appreciated until
the 80s when experimental methods were finally up to the task of
constructing molecular machines. One of Feynman's
challenges---to 
print the \emph{Encyclopædia Britannica} on the head of a
pin---was only cracked in 1985.

\marginnote{
  \begin{center}
    \includegraphics[width=0.75\linewidth]{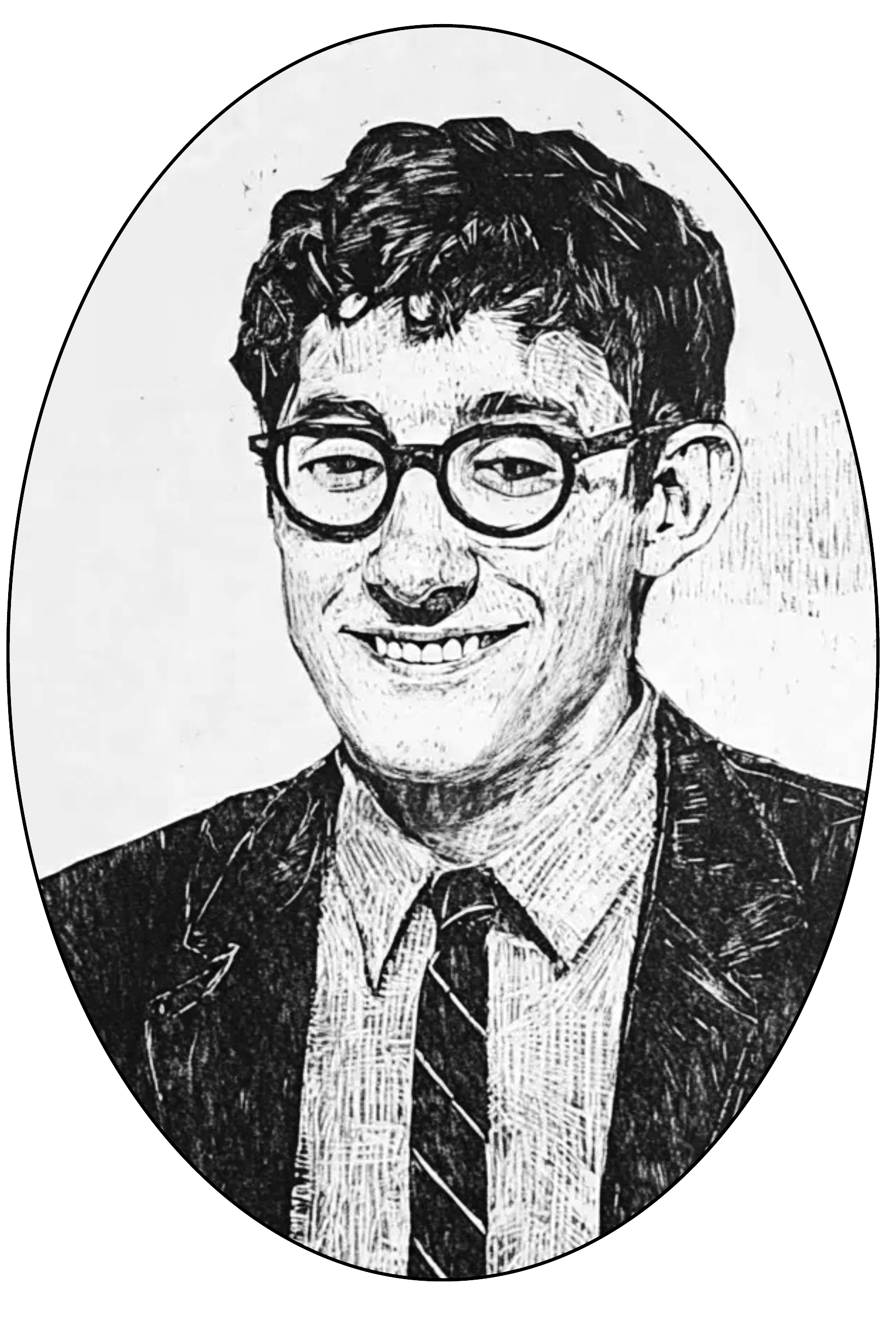}
  \end{center}
  \vspace{-10pt}
  \emph{Edward Fredkin (1934--2023). Millionaire, MIT professor
    and college dropout. Take from that what you will.
    \vspace{10pt}
  }
} While getting the \emph{Britannica} to dance on the head of a pin is a colourful
Feynman-esque conceit, more intriguing was his brief mention of
miniaturizing computers, fifteen years after his ENIAC tour of duty
and ten years before the first microprocessor:
\begin{quotation}
$\ldots$there is plenty of room to make [computers] smaller. There is nothing
that I can see in the physical laws that says the computer elements
cannot be made enormously smaller than they are now. In fact, there
may be certain advantages.
\end{quotation}These ``advantages'' were left mostly unspecified, and Feynman moved
on to other tasks---his Caltech lectures, the puzzles of
partons, a new Lagrangian-based approach to quantum mechanics---that
were more pressing and immediately soluble.

Feynman might never have returned to the problem were it not for a
college dropout called \textsc{Edward Fredkin}. Fredkin, a self-taught
programmer, floated between
consulting, industry, and sporadic faculty appointments at Carnegie Mellon and MIT. 
\marginnote{
  \begin{center}
    \includegraphics[width=0.75\linewidth]{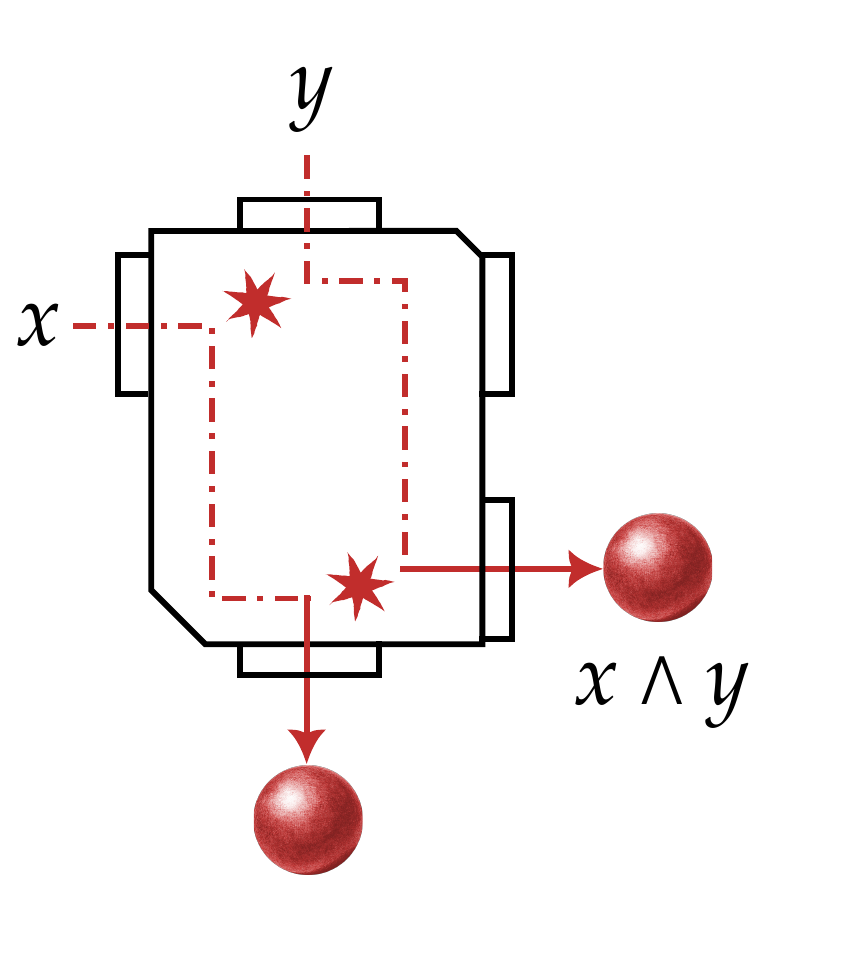}
  \end{center}
  \vspace{-15pt}
  \emph{\texttt{AND} gate for the reversible billiard-ball computer
    designed by Fredkin and Tomaso Toffoli. A ball represents
    $1$; its absence, $0$.
    \vspace{10pt}
  }
}One such stint was as Director
of Project MAC at MIT (a predecessor of CSAIL) from 1971--74;
after three years he got bored, and decided to head to Caltech to
spend time with Feynman, who he'd met in 1962 and found enjoyably
provocative. They struck a deal. Fredkin would stay for a year and teach Feynman
about computing; Feynman would teach him about quantum physics. Both
were somewhat skeptical about what was on offer, but committed to
learning.


It was a slow burn win-win. Fredkin successfully mastered quantum mechanics,
but was unconvinced the universe could be fundamentally continuous (it
didn't compute!). He tranformed that resistance into a successful research program for
``digital physics'', where discrete objects like cellular automata
were used to effectively mock up known physical laws.
It also motivated him to explore \emph{reversible} computation, since
all microscopic laws are time-reversal
invariant.\footnote{Technically, \textsf{CPT}-invariant, but we won't
  split hairs. See ``Conservative Logic'' (1982), Fredkin and Toffoli.} 
The burn was slower for Feynman.
He remained unsure that physics and
computation could be usefully connected;
maybe, after the bomb, he didn't want to connect them.

Regardless, the two remained close, and in 1981, Fredkin invited the
physicist to give the keynote at an MIT 
conference on physics and computation. It was going to be a lot of digital physics
``guff'' and Feynman was reluctant; he agreed, however, after Fredkin
gave him carte blanche on the topic. Feynman opened with this warm and
revealing tribute:\sidenote{``Simulating Physics with Computers''
  (1982). On the other side of the Iron Curtain,
  Yuri Manin independently had the same idea in 1980.}

\begin{quotation}
  The reason for doing this is something that I learned about from Ed
  Fredkin, and my entire interest in the subject has been inspired by
  him. It has to do with learning something about the possibilities of
  computers, and also something about possibilities in physics.
\end{quotation}
Feynman's hour-long address would explore the possibilities posed by simulating the physics
of the very small, propose a new type of machine called a \emph{quantum
computer} to address it, \marginnote{
  \vspace{2pt}
  \begin{center}
    \hspace{-10pt}\includegraphics[width=0.65\linewidth]{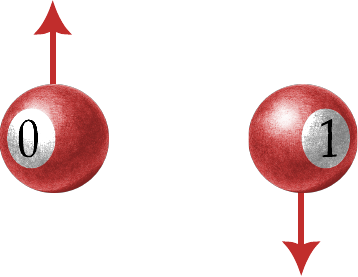}
  \end{center}
  \vspace{-3pt}
  \emph{Spin-$\tfrac{1}{2}$ particles have two
    states: \texttt{up} $|0\rangle$ and \texttt{down} $|1\rangle$. Feynman also
    used \texttt{absent}/\texttt{present}, like the billiard balls.
    \vspace{10pt}
  }
}and kickstart
a whole new field of computational science in the process.

Feynman's intuition was simple: it should be easier to imitate quantum
physics with a computer running on quantum principles.
He gave a heuristic argument to this effect, and outlined a
scheme for universal quantum simulation using what he called
spin-$\tfrac{1}{2}$ systems. These have two possible states,
usually denoted $|1\rangle$ and $|0\rangle$, so they are the quantum
analogue of a bit, also called a \emph{qubit}.
\marginnote{
  \vspace{2pt}
  \begin{center}
    \includegraphics[width=0.9\linewidth]{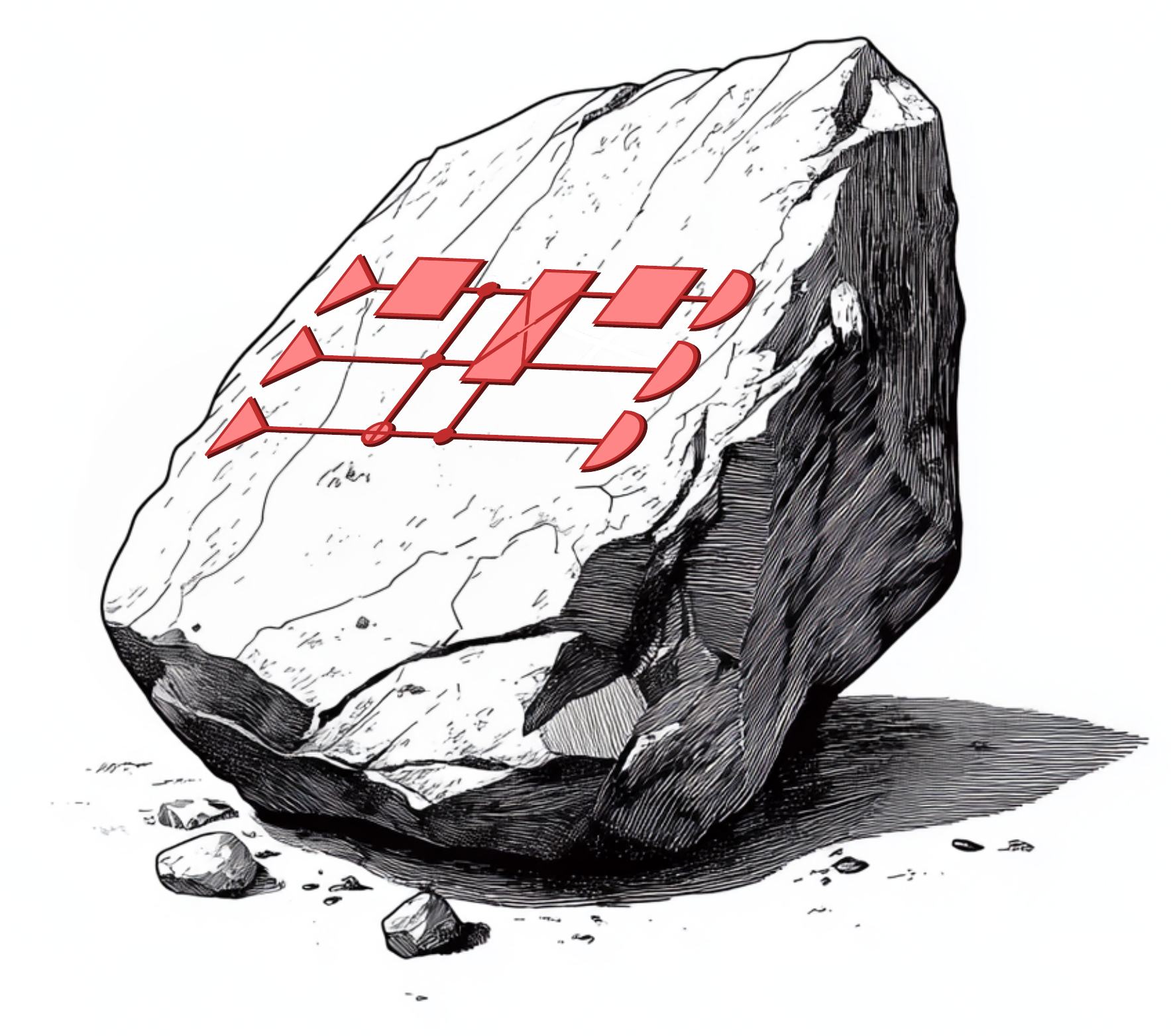}
  \end{center}
  \vspace{-3pt}
  \emph{Quantum binary in silico, aka quantum circuits. 
   Quantum circuits exist because small-scale simulation is hard.
    \vspace{10pt}
  }
}Though Feynman's motivations were rather different from the assembled
group, many
early contributors to quantum computing---Charles Bennett, Norman Margolus,
Tomaso Toffoli, and Fredkin himself---were present at the talk.
The way we reason about quantum computing, using qubits, circuits, and
reversible logic, bears their digital imprint.

It's tempting (and indeed customary) to view the qubit as the natural
endpoint of this tortuous back-and-forth between human and rock.
We scratched in unary on rocks, etched binary in metal, then
listened carefully and let metal teach us a new type of binary.
Now we are scaling Feynman's ladder in reverse and
extrapolating hardware from the very small to the macroscopic, hoping to perform
simulations more powerful than Feynman or von Neumann ever dreamed of.
It's a nice story, and it happens to be the reality we live in. But it didn't have to be.

\section[\emph{3} \hspace{5pt} A fork in the roadmap]{\LARGE{\emph{3}
    \hspace{5pt} A fork in the roadmap}}\label{sec:thinking}

We can try to picture a different path. It starts around 3000 light years away, with middle-aged couple danceing slowly 
through space. \textsc{T Coronae Borealis (T CrB)} is a binary system in the
constellation Coronae Borealis and the figurative jewel in its crown, consisting of a red giant and a
white dwarf gradually accreting material from its larger companion. Every 80
years or so, the white dwarf takes a giant slurp of charged matter and
blazes into view; this \emph{recurrent nova} briefly outshines most
stars in the sky. It blazes not only in radiation, but sometimes lone
protons, which are whipped around this stellar cyclotron and flung out
at close to the speed of light. We call these protons cosmic rays.
\marginnote{
    \vspace{-25pt}
  \begin{center}
    \includegraphics[width=0.75\linewidth]{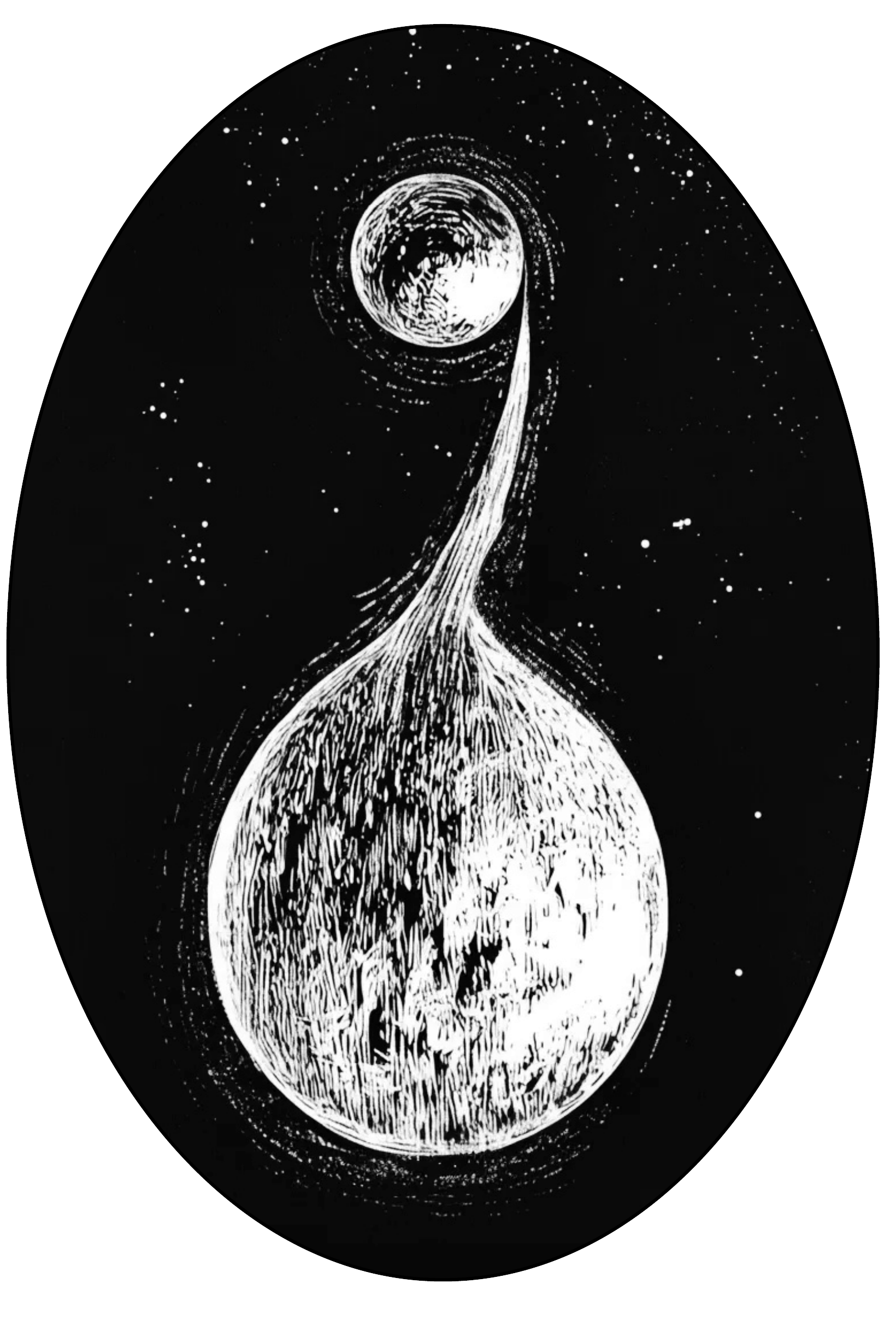}
  \end{center}
  \vspace{-10pt}
  \emph{T Coronae Borealis (1866--).
    The \emph{deus ex stellae} of our alternate timeline.
    \vspace{10pt}
  }
}

When cosmic rays hit the atmosphere, they fragment into a
cascade of secondary particles. 
This is a quantum process, described by Feynman diagrams; the effect of a cosmic
ray depends on the fragments.
In February 1946, T CrB went nova, with another luminosity bump in
June.
In some quantum fork of history, a cosmic ray broke up over
Philadelphia, showered a bank of vacuum tubes with ionizing radiation,
and knocked ENIAC
offline a month before it was due to be handed over to the military.
The operators knew it couldn't be background radiation---that struck
one tube at a time---and Mauchly and Eckert, now running the Electronic Control Company, began to speculate
about new kinds of systemic failure. An Ordnance Corps
concerned about their strategically crucial, multi-million dollar
investment requisitioned von Neumann from his efforts to design a new stored-program computer at the
IAS.\sidenote{See ``Preliminary discussion of the logical design of an electronic computing instrument'' (1946), Burks, Goldstine, and von Neumann.}

In July of 1946,
Operation Crossroads\sidenote{When history hands you a name this perfect, you don't refuse.}
would take place on Bikini Atoll, where some
physicists
would receive possibly lethal doses of radiation from the spectacular
but mismanaged Baker test.
\marginnote{
    \vspace{5pt}
  \begin{center}
    \includegraphics[width=\linewidth]{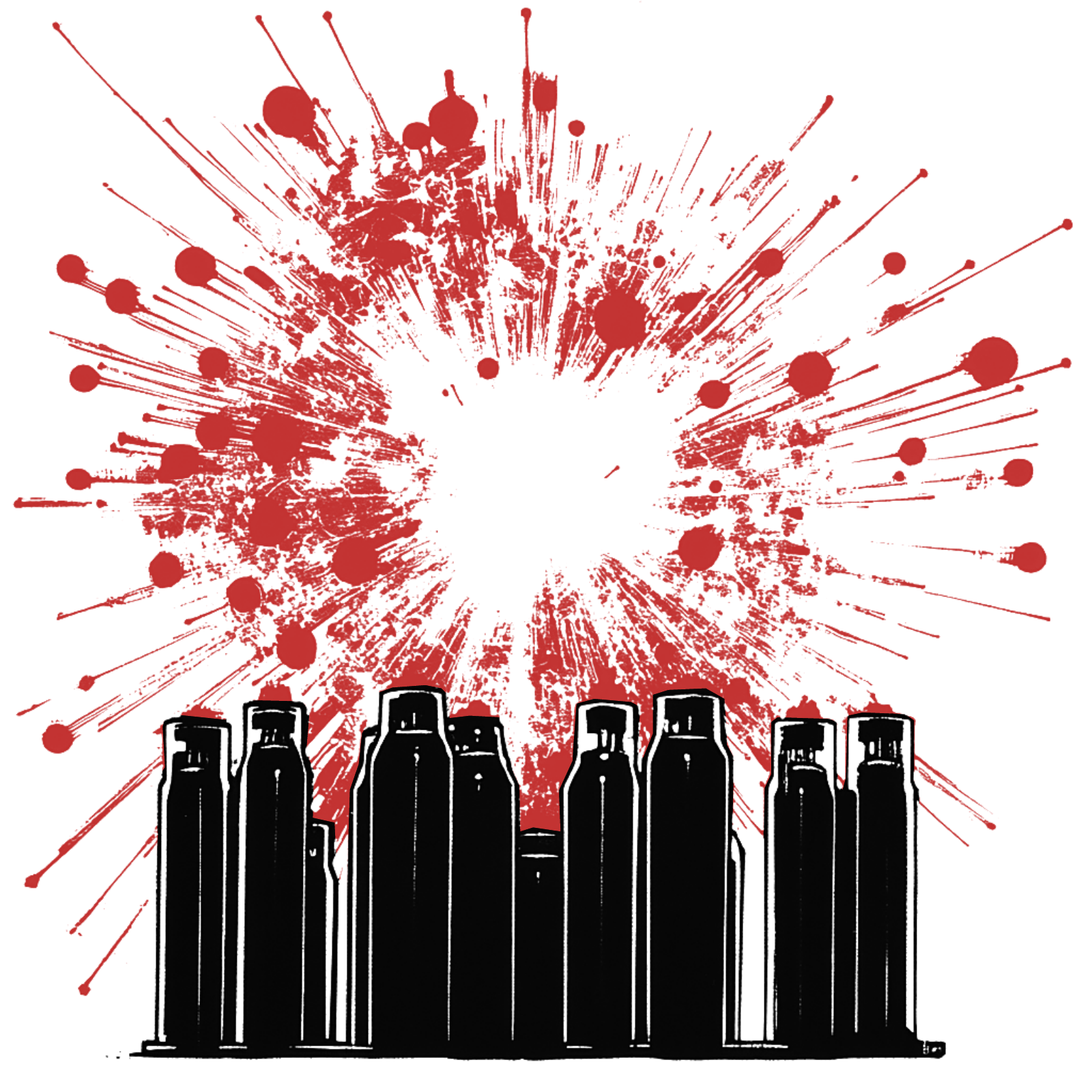}
  \end{center}
  \vspace{-7pt}
  \emph{The cosmic ray event at ENIAC that could have saved John von Neumann's life.
    \vspace{10pt}
  }
}In reality, von Neumann was one of these physicists; in the fork, he wasn't there. He was in Pennsylvania, where
ENIAC's flipped tubes
revealed a subtle directional gradient in ionization;
from the time of failure, the gradient pointed to a spot in the sky with right ascension $16$ hours and
declination $+26^\circ$. Then he was in Chicago, talking to
his old collaborator Subrahmanyan Chandrasekhar about radiative
transfer, then Pasadena to discuss novae with Walter Baade
and cosmic showers with his old boss Oppenheimer.
Finally, he flew back to New Jersey, where he argued with Stibitz
and Hamming at Bell Labs and stayed up late playing chess with Shannon.

Von Neumann had immediately guessed the culprit was a cosmic ray. The
astrophysicists confirmed this guess and offered a
candidate, T CrB; Oppenheimer, adrift at Caltech, was happy to tease
out the energetics of fragmentation. 
At Bell, Hamming and Stibitz
would talk guardedly about error correction, while Shannon, with
typical modesty, would outline a foundational perspective on noisy
channels.
Von Neumann took it all in, overlaid and interfered the
conversations, and by the time he returned to Princeton had arrived at
a novel conclusion: instead of 
\emph{correcting}
cosmic rays, perhaps they could \emph{control} them? 
After all, they had created ``artificial'' rays for implosion imaging.\sidenote{See ``Flash
  radiography with 24 GeV/$c$ protons'' (2011), C. Morris et al.}
Why not compute quantum with quantum?

In this version of reality, von Neumann was led to think about quantum
computing more than thirty years before Feynman.
Feynman
had a genius for the miniscule; naturally, he wanted to use computers
to understand his small friends better, and turned to the techniques---states, transition amplitudes,
diagrams, reversible circuits---he was familiar with. 
The result is quantum computing as we know it.
Von Neumann, in contrast, was a near scale-invariant scientist. He
studied stars and planetary-scale weather systems, macroscopic
architectures, and long before the war, laid the mathematical foundations
for quantum mechanics. His goals and techniques
spanned many more orders of magnitude. 

Let's start at the bottom. Von Neumann worked with \textsc{David Hilbert}\footnote{See ``Uber die Grundlagen der
  Quantenmechanik'' (1927), Hilbert and von Neumann.} to define what
we call \emph{Hilbert space}. This is a vector space
$\mathcal{H}$ over the complex numbers $\mathbbm{C}$, with an
inner product $\langle ,\cdot, \rangle$ and closed with respect
to the induced norm.\sidenote{Loosely
  speaking, ``closed'' means that any point
  we can approach arbitrarily closely is contained in
  $\mathcal{H}$. The induced norm is simply $\Vert x\Vert^2 = \langle
  x, x\rangle$.} A \emph{state} is a unit length vector in this space.
This is the
usual arena of quantum mechanics and quantum computing.
Ten years later, however, von Neumann had become skeptical of the Hilbert
space formalism, writing\sidenote{Quoted in ``Why John von Neumann did not Like the Hilbert Space Formalism of Quantum Mechanics
(and What he Liked Instead)'' (1996), Miklós Rédei.}\marginnote{
    \vspace{5pt}
  \begin{center}
    \includegraphics[width=0.8\linewidth]{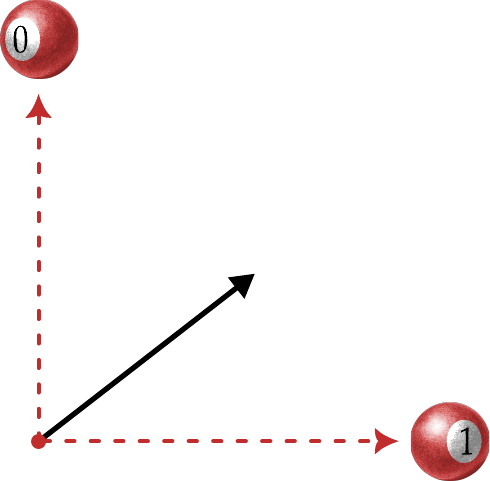}
  \end{center}
  \vspace{-02pt}
  \emph{Replacing states with projection operators, or equivalently,
    the $0$ and $1$ eigenspaces.
    \vspace{15pt}
  }
}
\begin{quotation}
  I would like to make a confession which may seem immoral: I do not
  believe absolutely in Hilbert space any more$\ldots$ Because: (1)
  The vectors ought to represent the physical states, but they do it
  redundantly, up to a complex factor, only (2) and besides, the states
  are merely a derived notion, the primitive (phenomenologically
  given) notion being the qualities which correspond to the linear
  closed subspaces$\ldots$
\end{quotation}
By point (1), he means that a state $|\psi\rangle \in \mathcal{H}$, and
a state $e^{i\theta}|\psi\rangle$ differing only by a phase, are
physically equivalent. Point (2) takes as
``phenomonologically given'' special operators $\Pi: \mathcal{H}\to
\mathcal{H}$ satisfying
\[
  \Pi^2 = \Pi, \quad \Pi^\dagger = \Pi.
\]
These are called \emph{projection operators}, and the range $\mathcal{V}_\Pi =
\Pi(\mathcal{H})$ of the operator is a closed subspace of $\mathcal{H}$.
More physically, these projections correspond to binary ``yes''/''no'' measurements, in the sense that the
eigenvalues are $0$ (``no'') and $1$ (``yes'').

This bears more than a little resemblance to classical bits and their
algebraic realization. Indeed, early in his career, von Neumann pioneered the
theory of \emph{operator
  algebras},\sidenote{``On Rings of Operators I/II'' (1936/7), Murray
  and von Neumann.}
where instead of studying the vectors transformed by operators like
$\Pi$, we study the algebraic structure of the operators themselves.
This led him organically to the logical aspects of the problem, and he collaborated with \textsc{Garret Birkhoff} on
the lattice-theoretic characterization of closed linear subspaces and
many related problems. As Birkhoff put it, von Neumann's
``brilliant mind blazed over lattice theory like a
meteor'';\sidenote{``Von Neumann and lattice theory'' (1958), Garret
  Birkhoff.} a recurrent nova might be a better analogy.
Ultimately, though, this ``projective quantum logic'' does not capture
the true logical power of the quantum, and can be efficiently simulated on a classical computer.\sidenote{``Nondeterministic testing of Sequential Quantum Logic propositions on a quantum
  computer'' (2005), Matt Leifer.}
But an expertise in operator theory and quantum logic would have been
fertile ground when the right seed came along.


If von Neumann skipped the Bikini Atoll tests to troubleshoot ENIAC, \marginnote{
  \begin{center}
      \vspace{-20pt}
    \includegraphics[width=0.75\linewidth]{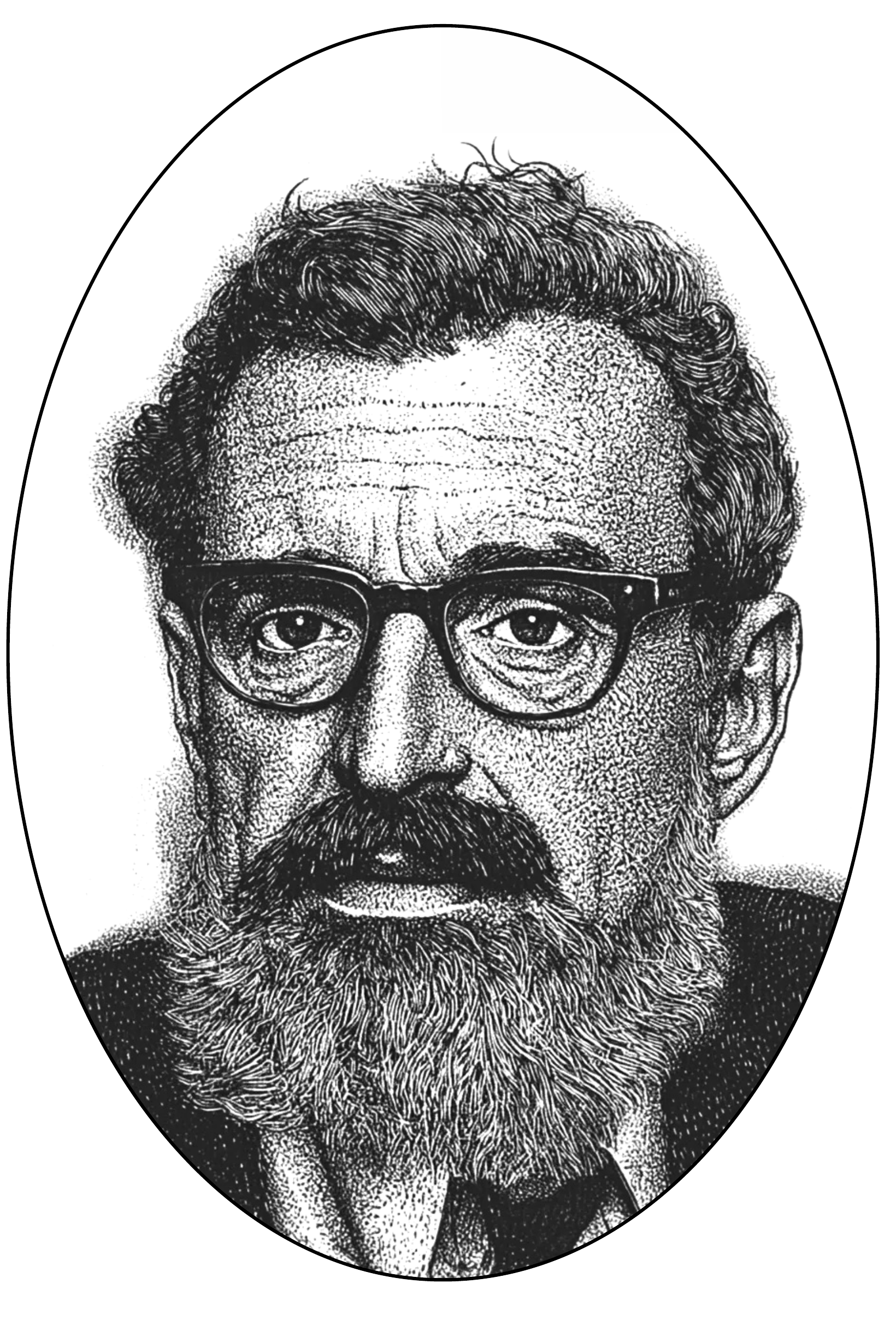}
  \end{center}
  \vspace{-15pt}
  \emph{Irving Ezra Segal (1918--1998). The mathematical prophet from the Bronx.
    \vspace{5pt}
  }
}it's
plausible that \textsc{Irving Segal}, a promising young 
mathematician freshly decommissioned from the Ballistic
Research Laboratory, 
would have heard about it.
Before the Laboratory, Segal had worked on operator algebras
with von Neumann at the IAS, where he had incidentally overlapped with Shannon.
He was planning to return northeast to winter at Princeton and continue thinking about
algebras and quantum mechanics, two topics
von Neumann had mostly abandoned in favour of building thermonuclear
weaponry.

Von Neumann's original work focused on the properties of the
projections $\Pi$ corresponding to binary measurements. In his view,
these operators would be the ``phenomenologically given''
analogue of bits, rather than states like $|0\rangle$ and $|1\rangle$ in
which he had ``lost faith''.
The algebra of operators generated by these binary measurements is
called a \emph{von Neumann algebra}.\sidenote{After his paper ``Zur Algebra der Funktionaloperationen und Theorie der
  normalen Operatoren'' (1930).} Like von Neumann, Segal wanted to
capture quantum mechanics algebraically, but he had a few new desiderata. First, physics allows for measurements
with richer outcomes than simply ``yes'' or ``no''.
He was especially worried by the ``mathematical difficulties in quantum electrodynamics,''\sidenote{
  ``Irreducible representations of operator
  algebras'' (1947).} which Feynman was in the process of
(non-rigorously) taming.
Second, not every operator on Hilbert space should be allowed;
we might not 
be able to measure it!
Segal needed a formalism more expressive than von
Neumann algebras, and more selective than Hilbert space.

He found it in the work of two Russian mathematicians,
Gelfand and Naimark,\sidenote{``On the embedding of normed rings into the
  ring of operators in Hilbert space'' (1943), I. Gelfand,
  M. Naimark.} who were trying to understand algebraic structures
called \emph{normed rings}, where you can add, multiply, and measure length.
They found a clever way to embed these rings as a set of operators on a
Hilbert space. To Gelfand and Naimark, this was merely a technical
bridge; to Segal, it was the royal road from algebra to quantum mechanics he had
been seeking.
The result of his meditations at Princeton was a paper \marginnote{
    \vspace{-20pt}
  \begin{center}
    \includegraphics[width=0.85\linewidth]{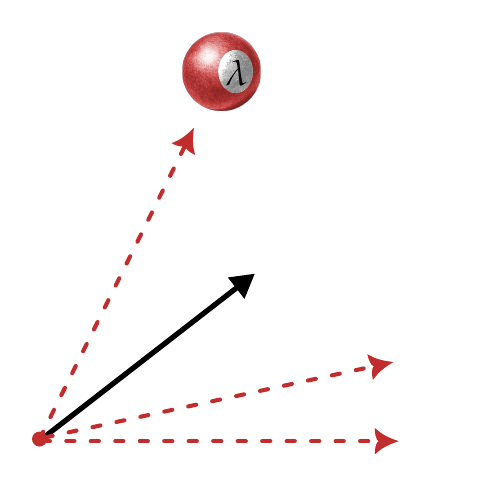}
  \end{center}
  \vspace{-4pt}
  \emph{In a general C*-algebra, we cannot reduce every measurement to
    its projections.
    \vspace{15pt}
  }
}that baptized the
normed rings
\emph{C*-algebras}, connected them to quantum mechanics, and improved
the embedding techniques. The method of
identifying an abstract C*-algebra with a concrete set of operators
on a
Hilbert space is called the \emph{Gelfand-Naimark-Segal (GNS) construction} in
their honour.

%

To see how this algebra meets Segal's criteria, note that in the usual
treatment of quantum mechanics, \emph{any} self-adjoint
linear operator $T = T^*$ on Hilbert space
is an observable.
A C*-algebra is much finer-grained; we include only the
observables we care about, along with the things we can generate from them.
A von Neumann algebra $\mathcal{A}$ also turns out to be a special type of
C*-algebra where, for any operator $T \in \mathcal{A}$ and eigenvalue
$\lambda$ of $T$, the projection $\Pi_{T, \lambda}$ onto the $\lambda$-eigenspace of $T$ is in
$\mathcal{A}$.
But a general C*-algebra permits quantum reality to be richer than
yes/no answers can exactly capture.

Segal would make a career out of deep, unexpected connections
between math and physics, helped in part by his stubborn and
unyielding individuality. As his AMS obituary
concludes,\footnote{``Irving Ezra Segal (1918–1998)'' in \emph{AMS
  Notices} (1999).}
\begin{quotation}
  $\ldots$ the full impact of the work of Irving Ezra Segal will become known only to future generations.
\end{quotation}
And in the words of John Baez:\sidenote{Ibid.}
\begin{quotation}
  Everyone who knows Segal will recall his inability to do things any way other than his own. 
\end{quotation}
A short kid from the Bronx, Segal learned early to stand up for
himself; a career in mathematical physics did not take the Bronx
out of the boy.
Despite his evident genius, von Neumann did not have Segal's depth,\sidenote{He was
  plagued by self-doubt, opining that he would be forgotten but ``Gödel remembered with Pythagoras.''}
though his breadth was unrivalled by any mathematician of the 20th
century, perhaps history.
If, by some cosmic glitch, Johnny was spared from the Baker test
and guided towards computing with the quantum,
deep might have converged with broad at Princeton, the winter of 1946, for a project of mutual interest.

\section[\emph{4} \hspace{5pt} Through the looking glass]{\LARGE{\emph{4}
    \hspace{5pt} Through the looking glass}}\label{sec:thinking}


\marginnote{
    \vspace{-20pt}
  \begin{center}
    \includegraphics[width=0.8\linewidth]{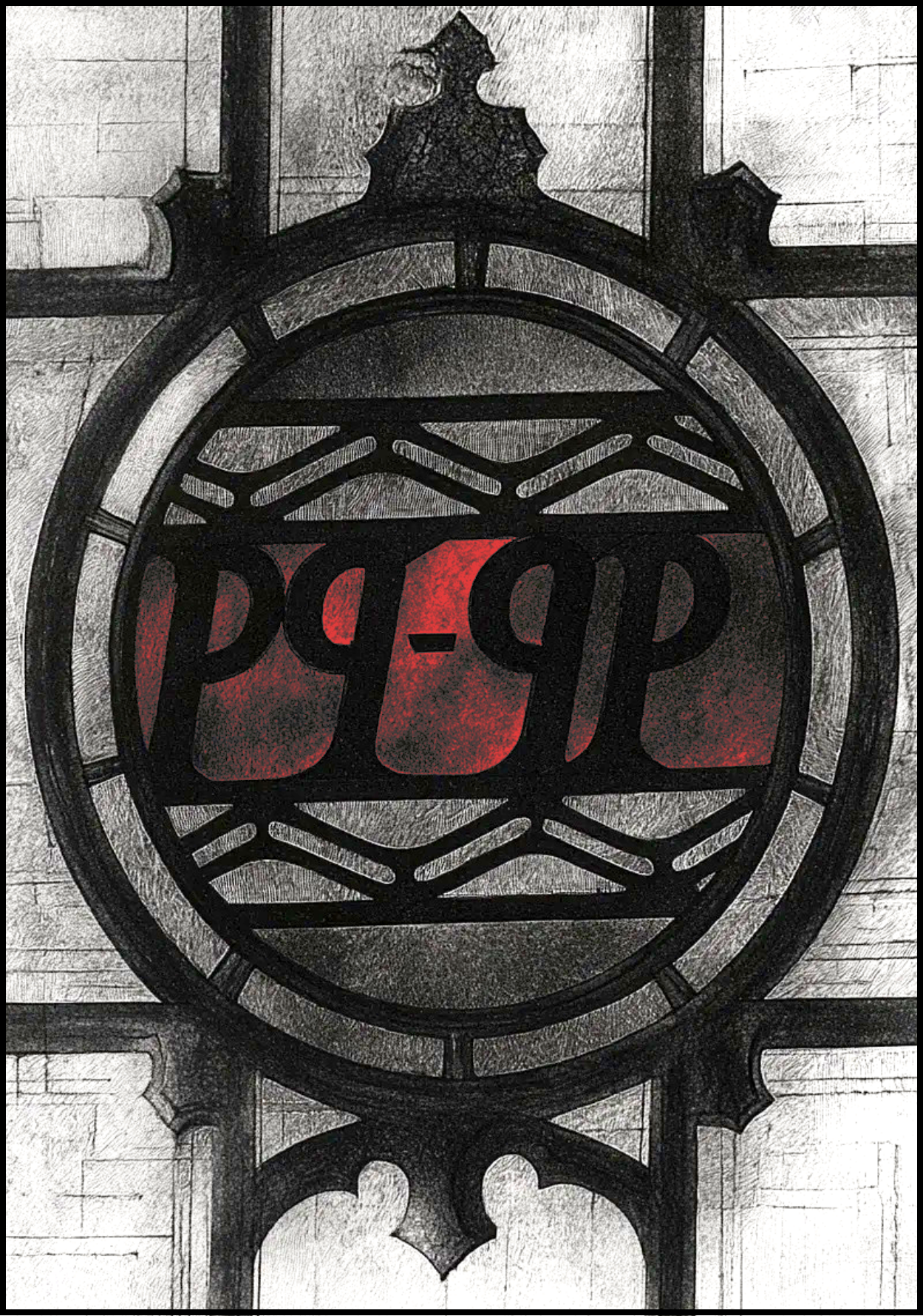}
  \end{center}
  \vspace{-4pt}
  \emph{The left-hand side of Heisenberg's uncertainty principle,
    emblazoned in leadlight.
    \vspace{15pt}
  }
}Imagine the two in the faculty lounge at Old Fine Hall, pulling
out a chalkboard to trade ideas; von Neumann with rapidfire cerebration,
Segal his Bronx one-two of prickle and boldness.
Behind them, the afternoon sun slants through stained-glass renderings
of the uncertainty principle and relativity.
Segal confides his goal of axiomatizing quantum mechanics with general
normed rings,
since, gesturing towards the windows, he suspects that ``a relativistic continuum may not admit
projections''.
Von Neumann asks some questions, then suggests in his offhand way a variant of 
Gelfand and Naimark's method of embedding the ring in Hilbert space,
entirely ignorant of their results. Segal is wryly shocked,
recovers, suggests improvements, and the two go on, developing
C*-algebras long into the evening.

They continue over the next few days, fleshing out the
representation theory and the rudiments of an axiomatic treatment of
quantum fields. Von Neumann encounters a few little roadblocks, gets bored,
then like the projectiles he has spent so much time on, launches himself
once more: heading over to engineering to blueprint designs
for an ENIAC successor, to Philadelphia to
talk shop with Mauchly and Eckhert,
and finally to Washington to sit on a nuclear advisory
committee and nag the Weather Bureau for more money.

%
But the algebra won't leave him alone.
Von Neumann goes to sleep one night thinking
about ENIAC, programmability and the structure of
cosmic rays$\ldots$ 
The next day, he drives in a state of cheerful mania to Manhattan.
The magician of Bell Labs receives him with courteous surprise, and
they walk to a park 
overlooking the Hudson:\sidenote{This dialogue, like the interaction with Segal, is fictitious.}
\begin{quotation}
  \textsc{jvn}: So, Shannon, you know the law of idempotency in a Boolean algebra,
  $x^2 = x$. [Shannon nods.] These variables are also
  commutative, in the sense that $x\cdot y = y \cdot x$, so your
  $\texttt{AND}$ connective doesn't care about order of conjuncts. \marginnote{
  \vspace{-0pt}
  \begin{center}
    \includegraphics[width=0.9\linewidth]{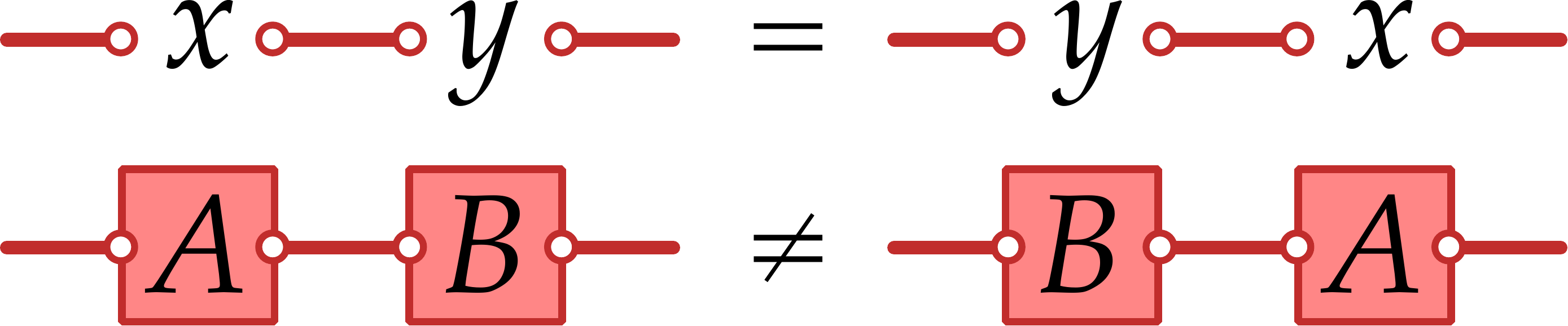}
  \end{center}
  \vspace{3pt}
  \emph{Above, a commutative circuit where switches, corresponding to
    Boolean variables, can be interchanged. Below, a noncommutative circuit where
    switches are C*-algebraic variables which cannot always be interchanged.
    \vspace{10pt}
  }
}There's a
  noncommutative version of this where we replace Boolean variables
  by operators, and in particular we could build our theory around projectors obeying $\Pi^2 = \Pi$,
  so-called because they project vectors onto a subspace. 
  [Shannon pauses briefly, then nods.] Good.
  Now, remember Irving, who was always disagreeing with Einstein? [A smile.] Well, he
  found a clever way to take rings of operators defined
  abstractly---like your Boolean algebra---and 
  represent them as transformations of a 
  Hilbert space. I want to
  understand if we can complete the parallelogram and build
  noncommutative circuits.
  \\
\vspace{5pt}
  \noindent \textsc{ces}: I see. [Pause.] There is a perfect correspondence
  between the terms of a Boolean formula and the structure of the
  relay or switching
  circuit because, after a suitable identification of the components, physical
  laws of combination precisely match symbolic ones. \\
  \vspace{5pt}
  \noindent \textsc{jvn}: Yes yes, we need our algebraic laws to map
  onto the noncommutative laws obeyed by our wires. Enter
  quantum theory. Guided by my ancient work on orthocomplemented lattices, I have
  this [fishes napkin out of suit pocket] mildly soiled design schematic for
  inducing transitions in caesium$\ldots$ But what makes the circuits
  morally necessary?
  \\
  \vspace{5pt}
  \noindent \textsc{ces}: In what you call the
  ``commutative'' case, the hindrance of a complex
  circuit is determined by the hindrance of its components. [Long
  pause.] Hindrance means the passage of current, so I suppose we can
  view this as a law by which aspects of the physical state of the circuit are
  determined by the state of its components.
    \\
  \vspace{5pt}
  \noindent \textsc{jvn}: You mean ``state''
  colloquially, but it's also a mathematical term of art for
  assigning
  numbers to operators, like current.
  \\
    \vspace{5pt}
    \noindent \textsc{ces}: Yes. I suppose a ``state'' here
    is a function assigning binary values to each independent
    propositional variable. Equivalently, this is an entry in a truth
    table. [Pause.] How is Segal's construction related?
    \\
    \vspace{5pt}
    \noindent \textsc{jvn}: [Manic grin.] Ah, Shannon. Ah! The states of
    Segal's embedding are lines in the truth table. Irreducibility
    means restriction to the involved variables$\ldots$
    But all this is trivial. Now, the question I pose to you, say
    Stibitz, Hamming of course, maybe
    Bardeen: can we build it?
\end{quotation}
\marginnote{
  \vspace{-60pt}
  \begin{center}
    \includegraphics[width=0.9\linewidth]{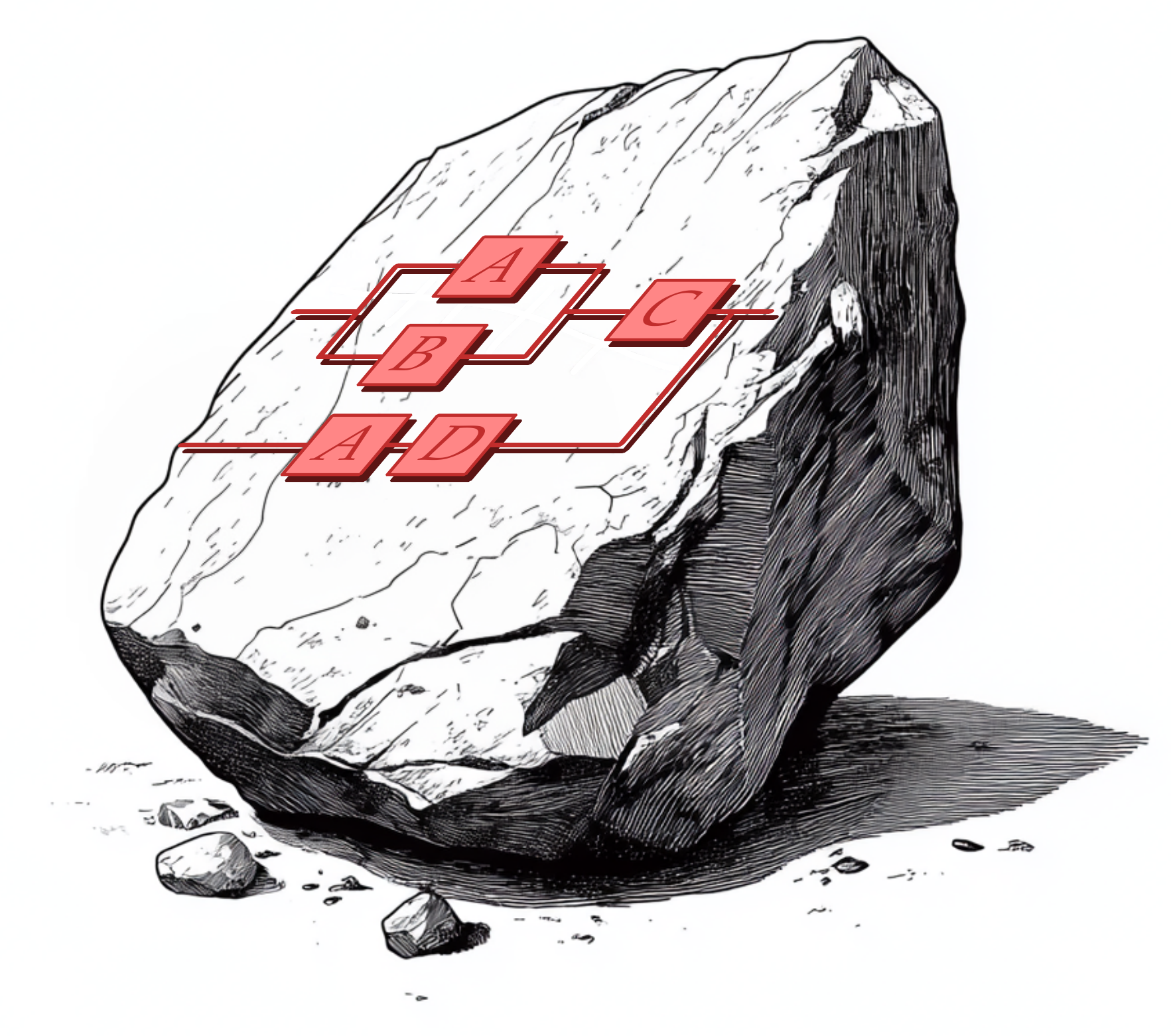}
  \end{center}
  \vspace{-3pt}
  \emph{C*-algebraic programming in silico, aka noncommutative circuits.
   These circuits do not exist, but might have if a cosmic ray struck ENIAC.
    \vspace{10pt}
  }
}Von Neumann leaves West Village with visions of ``noncommutative
architectures,'' algebras as data and states as programs, his mind
racing into the future. Shannon 
mentions the conversation to Mervin Kelly, director of
research at Bell Labs, who thinks it over, heads to Princeton, and
on his return quietly commissions a new research program.
In branch T CrB, the rocks instructed us to replace hindrance with expectation, truth-tables with GNS, and Boolean algebras with functional analysis.
This is
not the branch we live in. But with a little imagination and some
mathematical elbow grease, we can catch up.

\section[\emph{5} \hspace{5pt} Postscript]{\LARGE{\emph{5}
    \hspace{5pt} A Westbeth postscript}}\label{sec:thinking}

\marginnote{
  \begin{center}
      \vspace{0pt}
    \includegraphics[width=0.75\linewidth]{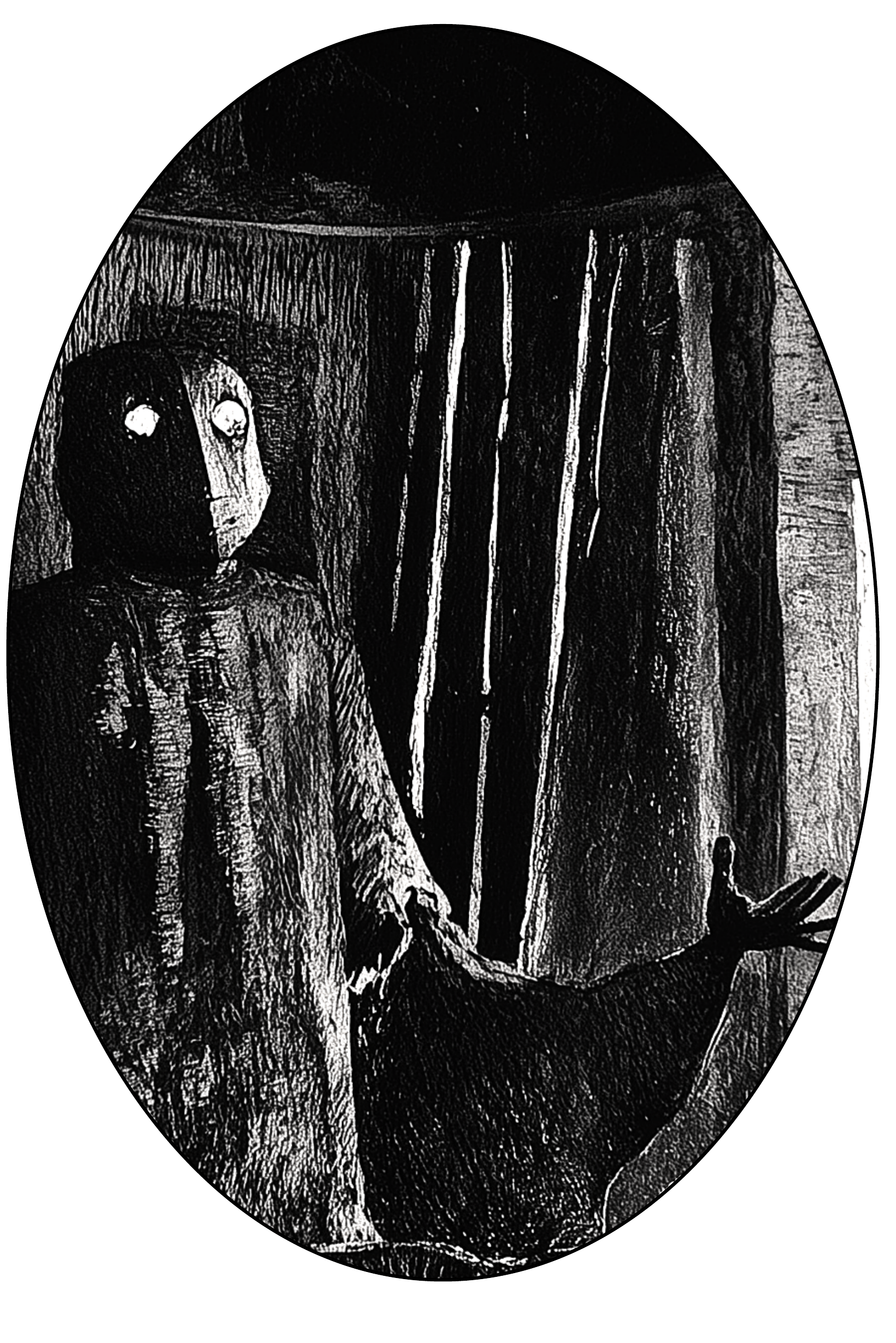}
  \end{center}
  \vspace{-5pt}
  \emph{The mad puppets of Ralph Lee (1936--2023), oblique testament to
    the spirit of Bell Labs.
    \vspace{5pt}
  }
}\noindent Visiting the old Bell Laboratories Building at $463$ West
Street, it's hard to detect the \emph{genius loci} that moved it in
former times. In the 60s, Bell moved its operations from Manhattan to a shiny new facility in
New Jersey, and the West Street complex was converted into Westbeth
Artists Community, the largest artist cooperative in the
world. Its roster of luminaries includes photographer Diane Arbus, the
choreographer Merce
Cunningham, and puppeteer Ralph Lee, whose shambolic creatures
still grace the commune in unexpected corners.
It is more Coney Island than Wall Street. 

After hours, I snuck in behind a departing resident and
found myself in an interstitial maze of
shoebox consultancies, potters' studios, office space,
and miscellaneous storage, a \emph{wunderkammer} of uninterpretable objects$\ldots$
I turned around and left before I could get irrevocably lost.
Back in the courtyard, I understood that the spirit of Bell Labs
remained alive
in Westbeth:
in its organized shaping of creative energies, its zest for
human excellence,
in scale and the network effects that come from so many unusual minds
melting in the same pot.
Shannon made theories in the same way that Lee made puppets: from an urge
to make something beautiful, strange and new in the world, and from
that urge alone.

The bricks have many stories to tell; Westbeth listens.
Standing in the courtyard, I too strain to hear, and erect the
puppets of von Neumann and Shannon a hundred yards away in the process.
But there the whispers die; the rocks can only tell us so much.
The work of building a noncommutative \emph{ratiocinator}, the
high-level \emph{characteristica} to go with it, and a broader
institutional culture of rock-listening---completing the arc of that
proton over Philadelphia---is left to us.

\nocite{*}

  \bibliography{fqp}

\begin{thebibliography}{10}

\bibitem{AMSNotices1999}
{\sc Baez, J.~C., Beschler, E.~F., Gross, L., Kostant, B., Nelson, E., Vergne,
  M., and Wightman, A.~S.}
\newblock Irving {E}zra {S}egal (1918–1998).
\newblock In {\em Notices of the American Mathematical Society}, A.~M. Society,
  Ed., vol.~46. American Mathematical Society, 1999.

\bibitem{Birkhoff:1958:NLT}
{\sc Birkhoff, G.}
\newblock {Von Neumann} and lattice theory.
\newblock {\em Bulletin of the American Mathematical Society 64}, 3 (May 1958),
  50--56.

\bibitem{Blair}
{\sc {Blair, Jr.}, C.}
\newblock Passing of a great mind.
\newblock {\em Fortune\/} (1957).

\bibitem{Boole_2009b}
{\sc Boole, G.}
\newblock {\em An Investigation of the Laws of Thought: On Which Are Founded
  the Mathematical Theories of Logic and Probabilities}.
\newblock Cambridge University Press, 2009.

\bibitem{Boole_2009a}
{\sc Boole, G.}
\newblock {\em The Mathematical Analysis of Logic: Being an Essay Towards a
  Calculus of Deductive Reasoning}.
\newblock Cambridge University Press, 2009.

\bibitem{Burks}
{\sc Burks, A.~W., Goldstine, H.~H., and von Neumann, J.}
\newblock Preliminary discussion of the logical design of an electronic
  computing instrument, 1946.

\bibitem{3ca88a47-e774-3e1e-97ed-ea58ceb26f65}
{\sc Carson, C.~L.}
\newblock An {E}den after the {F}all.
\newblock {\em Reviews in American History 21}, 3 (1993), 514--519.

\bibitem{Charney:1950:NIB}
{\sc Charney, J.~G., Fj{\"o}rtoft, R., and von Neumann, J.}
\newblock Numerical integration of the barotropic vorticity equation.
\newblock {\em Tellus 2\/} (1950), 237--254.

\bibitem{feynman1992surely}
{\sc Feynman, R., Leighton, R., and Hutchings, E.}
\newblock {\em "Surely You're Joking, Mr. Feynman!": Adventures of a Curious
  Character}.
\newblock Vintage, 1992.

\bibitem{Feynman}
{\sc Feynman, R.~P.}
\newblock There's plenty of room at the bottom: An invitation to enter a new
  field of physics.
\newblock In {\em Miniaturization}, H.~D. Gilbert, Ed. Reinhold, 1961.

\bibitem{feynman1982simulating}
{\sc Feynman, R.~P.}
\newblock Simulating physics with computers.
\newblock {\em International {J}ournal of {T}heoretical {P}hysics 21}, 6/7
  (1982), 467--488.

\bibitem{FT:82b}
{\sc Fredkin, E.~F., and Toffoli, T.}
\newblock Conservative logic.
\newblock {\em International {J}ournal of {T}heoretical {P}hysics 21}, 3/4
  (1982), 219--253.

\bibitem{Гельфанд1943}
{\sc Gelfand, I., and Naimark, M.}
\newblock On the imbedding of normed rings into the ring of operators in
  hilbert space.
\newblock {\em Sbornik Mathematics 54}, 2 (1943), 197--217.

\bibitem{6106}
{\sc Gertner, J.}
\newblock {\em The {I}dea {F}actory: {B}ell Labs and the great age of American
  innovation}.
\newblock Penguin Books, New York, 2013.

\bibitem{Hilbert1928}
{\sc Hilbert, D., von Neumann, J., and Nordheim, L.}
\newblock Über die grundlagen der quantenmechanik.
\newblock {\em Mathematische Annalen 98\/} (1928), 1--30.

\bibitem{Leibniz1989b}
{\sc Leibniz, G.~W.}
\newblock {\em Letters To Nicolas Remond}.
\newblock Springer Netherlands, Dordrecht, 1989, pp.~654--660.

\bibitem{Leibniz1989a}
{\sc Leibniz, G.~W.}
\newblock {\em On the General Characteristic}.
\newblock Springer Netherlands, Dordrecht, 1989, pp.~221--228.

\bibitem{leifer2005nondeterministictestingsequentialquantum}
{\sc Leifer, M.~S.}
\newblock Nondeterministic testing of sequential quantum logic propositions on
  a quantum computer, 2005.

\bibitem{10.1063/1.3580262}
{\sc Morris, C.~L., Ables, E., Alrick, K.~R., Aufderheide, M.~B., Barnes,
  P.~D., J., Buescher, K.~L., Cagliostro, D.~J., Clark, D.~A., Clark, D.~J.,
  Espinoza, C.~J., Ferm, E.~N., Gallegos, R.~A., Gardner, S.~D., Gomez, J.~J.,
  Greene, G.~A., Hanson, A., Hartouni, E.~P., Hogan, G.~E., King, N. S.~P.,
  Kwiatkowski, K., Liljestrand, R.~P., Mariam, F.~G., Merrill, F.~E., Morgan,
  D.~V., Morley, K.~B., Mottershead, C.~T., Murray, M.~M., Pazuchanics, P.~D.,
  Pearson, J.~E., Sarracino, J.~S., Saunders, A., Scaduto, J., Schach~von
  Wittenau, A.~E., Soltz, R.~A., Sterbenz, S., Thompson, R.~T., Vixie, K.,
  Wilke, M.~D., Wright, D.~M., and Zumbro, J.~D.}
\newblock Flash radiography with 24 {G}e{V}/c protons.
\newblock {\em Journal of Applied Physics 109}, 10 (05 2011), 104905.

\bibitem{vonNeumann:1936:RO}
{\sc Murray, F.~J., and von Neumann, J.}
\newblock On rings of operators.
\newblock {\em Bulletin of the American Mathematical Society 42\/} (1936).

\bibitem{Murray:1937:ROI}
{\sc Murray, F.~J., and von Neumann, J.}
\newblock On rings of operators ({II}).
\newblock {\em Transactions of the American Mathematical Society 41}, 2 (1937),
  208--248.

\bibitem{REDEI1996493}
{\sc Rédei, M.}
\newblock Why {J}ohn von {N}eumann did not like the {H}ilbert space formalism
  of quantum mechanics (and what he liked instead).
\newblock {\em Studies in History and Philosophy of Modern Physics 27}, 4
  (1996), 493--510.

\bibitem{bams/1183510397}
{\sc Segal, I.~E.}
\newblock Irreducible representations of operator algebras.
\newblock {\em Bulletin of the American Mathematical Society 53}, 2 (1947), 73
  -- 88.

\bibitem{crypto}
{\sc Shannon, C.~E.}
\newblock A mathematical theory of cryptography, 1945.

\bibitem{shannon}
{\sc Shannon, C.~E.}
\newblock A mathematical theory of communication.
\newblock {\em The Bell System Technical Journal 27\/} (1948), 379--423.

\bibitem{energy-info}
{\sc Tribus, M., and McIrvine, E.~C.}
\newblock Energy and information.
\newblock {\em Scientific American 225}, 3 (1971), 179--190.

\bibitem{vonNeumann:1929:AFT}
{\sc von Neumann, J.}
\newblock Zur algebra der funktionaloperatoren und theorie der normalen
  operatoren.
\newblock {\em Mathematische Annalen 102\/} (1929), 370--427.

\end{thebibliography}
  \bibliographystyle{acm}

  \vspace{0pt}
  
\begin{flushleft}
  \textsc{colophon}
\end{flushleft}

\begin{abstract}
  This document is typeset using the
  \href{https://tufte-latex.github.io/tufte-latex/}{Tufte-\LaTeX}
  document class, with \emph{Palatino} as the body font,
  \href{https://fonts.google.com/specimen/IBM+Plex+Mono}{\texttt{IBM Plex
      Mono}} for teletype, and $AMS\, Euler$ for math.
  The primary visual inspiration was the
  \href{https://en.wikipedia.org/wiki/Life_Nature_Library}{Life Nature
    Library}, volumes of which can be found in any reputable thrift
  shop.
  Illustrations were create with a combination of Midjourney, Inkscape,
  and public domain images. 
  Finally, this is distributed under a
  \href{https://creativecommons.org/licenses/by-nc-nd/4.0/deed.en}{CC
    BY-NC-ND} license; feel free to redistribute in its current form, but if you wish to modify
  or use specific parts, please ask me directly.
\end{abstract}
  
\end{document}